\documentclass[%
 preprint,
 amsmath,amssymb,
 aps,
]{revtex4-2}

\usepackage{graphicx}
\usepackage{dcolumn}
\usepackage{bm}
\usepackage{epstopdf, epsfig}
\usepackage{psfrag}
\usepackage{color}
\newcommand{\tr}[1]{\textcolor{black}{#1}} 
\newcommand{\fc}[1]{\textcolor{black}{#1}} 
\newcommand{\rv}[1]{\textcolor{black}{#1}} 
\usepackage[mathlines]{lineno}

\begin{document}

\preprint{APS/Physical Review Fluids}

\title{Nonlinear aeroacoustic response of a harmonically forced\\ \tr{side branch aperture} under turbulent grazing flow: \\ modelling and experiments}

\author{Tiemo Pedergnana}
  \email{ptiemo@ethz.ch}
\author{Claire Bourquard}%
\author{Abel Faure-Beaulieu}%
\author{Nicolas Noiray}%
  \email{noirayn@ethz.ch}
\affiliation{\tr{CAPS Laboratory, Department of Mechanical and Process Engineering},  ETH Z{\"u}rich, Sonneggstrasse 3,
8092 Z{\"u}rich, Switzerland}%

\date{\today}

\begin{abstract}
Hydrodynamic modes in the turbulent mixing layer over a cavity can constructively interact with the acoustic modes of that cavity and lead to aeroacoustic instabilities. The resulting limit cycles can cause undesired structural vibrations \fc{or noise pollution} in many industrial applications. To further the predictive understanding of this phenomenon, we propose two \tr{physics-based} models which describe the nonlinear aeroacoustic response of a side branch \tr{aperture} under harmonic forcing with variable acoustic pressure forcing amplitude $p_a$. One model is based on Howe's classic formulation that describes the shear layer as a thin vortex sheet, and the other on an assumed vertical velocity profile in the side branch aperture. These models are validated against experimental data. Particle image velocimetry (PIV) was performed to quantify the turbulent and coherent fluctuations of the shear layer under increasing $p_a$. The specific acoustic impedance $Z$ of the \tr{aperture} was acquired \tr{over} a range of frequencies \tr{for different bulk flow velocities $U$} and \tr{acoustic pressure} forcing amplitudes \tr{$p_a$}.

In this work, we show that, once the handful of parameters in the two models for $Z$ have been calibrated using experimental data at a given condition, it is possible to \tr{make robust analytical predictions of} this impedance \tr{over a broad range of the frequency, bulk flow velocity and forcing amplitude.} In particular, the models allow prediction of a necessary condition for instability, implied by negative values of the acoustic resistance $\mathrm{Re}{(Z)}$\tr{, which corresponds to a reflection coefficient $R$ of the aperture with magnitude larger than $1$}. \tr{Furthermore, we demonstrate that the models} are able to describe the \tr{nonlinear saturation of the aeroacoustic response} caused by \fc{alteration} of the mean flow at large forcing amplitudes, which was recently reported in literature. This effect stabilizes the coupling between the side branch opening and the acoustic field in the cavity, and its quantitative description may be of value for control of aeroacoustic instabilities. We visualize and compare the models' representations of the \tr{hydrodynamic response in the side branch aperture and of the saturation effect} under increasing $p_a$.
\end{abstract}

\maketitle


\section{Introduction \label{sec:: Section 1: Introduction}}
Sound production through aeroacoustic instabilities that arise from the constructive interaction between acoustic modes of a cavity and the hydrodynamic response of a shear layer over that cavity is a classic and long-observed phenomenon in physics, which was first described in modern terms by \citet{sondhauss1854ueber}. Such instabilities occur, for example, when we whistle \cite{wilson1971experiments}, or play wind instruments, such as the \tr{ocarina or the organ pipe} \cite{fabre2012aeroacoustics}. The same feedback mechanism that is underlying these artistic applications leads, on a larger scale, to aeroacoustic instabilities in industrial machines which can cause noise pollution and fatigue damage of components \cite{rockwell1979self}. Practical aspects of the \tr{self-induced oscillation instability} mechanism and possible mitigation strategies are explored in \cite{ziada2014flow}. 

In the context of the classification of \citet{rockwell1978review}, the phenomena described above belong to the category of fluid-resonant cavity oscillations. This type of instability can be further subdivided into self-sustained oscillations of shallow cavities \cite{rowley2006dynamics}, which are governed by the mechanism described by \citet{rossiter1964wind}, and those of deep cavities, which are treated in detail, e.g., in the exhaustive work of \citet{howe_1998}. Regarding the latter group, we mention the pioneering research of \citet{elder1978self}, who, for a single deep cavity, analysed a feedback loop which incorporates the cavity opening and the aerodynamic forcing as a $\it{forward}$ transfer function and the acoustic resonance of the cavity as a $\it{backward}$ transfer function. \tr{In Elder's study, the forward transfer function, defined as a volume-flux gain, is derived from an estimated expression for the coherent, i.e., \rv{in phase with} the harmonic acoustic forcing, velocity fluctuations in the cavity aperture. This qualitative approach was adopted in \cite{MARSDEN20123521} to compute the forward transfer function of a round cavity opening, which is then used to study the interaction betwen the \fc{shear} layer and the acoustic field in the cavity. The present investigation is similar to those works in that we \fc{deduce} the aeroacoustic response of the \tr{side branch} opening \fc{from} the coherent velocity fluctuations in the aperture. The main difference is that we focus only on the acoustic impedance of the \tr{side branch} $\it{opening}$ and \fc{on the prediction of} its frequency-domain distribution as a function of the mean flow speed. Other notable differences are that\fc{,} instead of estimating the coherent velocity fluctuations, we obtain them indirectly from a parameter fit of the acoustic impedance to experimental data, and that we quantify the effect of large forcing amplitudes on the aeroacoustic response of the aperture.} \rv{Our approach bears some similarity to the work of \citet{YANG2016294}, who derive an analytical model of the acoustic impedance of an orifice under bias flow which is then compared to experiments. }

One can \fc{model} the aeroacoustic response of a side branch \tr{aperture} using various computational methods. We mention the work of \citet{martinez2009identification}, \tr{who combine incompressible flow simulations, vortex sound theory and system identification techniques to numerically compute the response at low Mach numbers.} Another approach is taken by \tr{\citet{Gikadi2012}, who use the compressible Navier--Stokes equations, linearized around a mean grazing flow obtained from large-eddy-simulations (LES).} They successfully compare the obtained transfer matrices to the experiments of \citet{karlsson2010aeroacoustics}.  \tr{In recently published studies,  \citet{fabre_longobardi_bonnefis_luchini_2019,fabre_longobardi_citro_luchini_2020} compute the acoustic impedance of a cirular aperture using LES simulations. In \cite{fabre_longobardi_bonnefis_luchini_2019}, this is achieved for a thin and in \cite{fabre_longobardi_citro_luchini_2020} for a thick wall.} 
\tr{Compressible} LES, combined with finite-element simulations of the linearized incompressible Navier--Stokes equations (LNSE)\tr{,} were also used by \citet{boujo_bauerheim_noiray_2018} to analyze the response of an acoustically forced \tr{side branch opening subject to a mean grazing flow with a bulk velocity of $56$ m/s.} In their setting, the mean flow, obtained from \tr{compressible} LES, is forced by a harmonic \tr{modulation of the velocity at the inlet of the side branch. The forcing frequency was set near the eigenfrequency of the main hydrodynamic mode, computed with LNSE analysis around the unforced LES mean flow.} The amplification of the forcing by the shear layer was studied using different \tr{quantitative} measures. Numerically, a \tr{nonlinear saturation of this amplification} was observed as the forcing velocity amplitude is increased, which is due to a thickening of the turbulent shear layer in the side branch aperture. This saturation effect leads to a decrease of the gain from the bulk flow, and its quantitative description remains a significant challenge for developing accurate predictive models of the aeroacoustic response of a side branch opening in the high-amplitude regime. \rv{We also mention the recent study of \citet{Bauerheim20}, who use LES to investigate the nonlinear (amplitude-dependent) vortex-sound interaction in a deep cavity and the hydrodynamics of the shear layer under increasing acoustic velocity amplitude. They find that, at small amplitudes, a flapping of the shear layer is responsible for the vortex sound generation. At high amplitudes, the acoustic response of the aperture saturates, roll-up of the shear layer occurs and shedding of discrete vortices is observed.} 

\tr{The saturation observed in the simulations of \citet{boujo_bauerheim_noiray_2018} is in agreement with the experiments of \citet{bourquard_faure-beaulieu_noiray_2021} that were performed with a square wind tunnel of the same height $H=62$ mm and with a side branch of the same width $W=30$ mm} at a mean bulk flow velocity of $74$ m/s, which corresponds to a Reynolds number $Re=UH/\nu$ of about $306\hspace{0.7mm}000$, where $\nu=1.5\times10^{-5}$ m$^2$/s is the kinematic viscosity of air. \tr{\fc{In \cite{bourquard_faure-beaulieu_noiray_2021},} the shear layer was forced \fc{over a broad frequency range in order to explain self-sustained aeroacoustic oscillations occurring for a closed side-branch and involving the three-quarter wave acoustic mode of the resulting deep cavity.}} The turbulent and coherent \tr{velocity} fluctuations of the shear layer in the center plane of the channel were extracted using PIV and the acoustic pressure signal. Using the multi-microphone method \cite{schuermans2004detailed}, the aeroacoustic response of the \tr{side branch opening} was measured for different bulk flow velocities $U$ and \tr{acoustic pressure forcing amplitudes} $p_a$ \tr{\fc{in the form of its} specific acoustic impedance $Z$, which links acoustic velocity and pressure at the opening.} \rv{For detailed information about the experimental set-up, the reader is referred to $\S$2 of \cite{bourquard_faure-beaulieu_noiray_2021}.} As in \cite{boujo_bauerheim_noiray_2018}, the acoustic forcing was imposed from the back of the side branch. They observed that for a certain \rv{values} of $U$, the measured \tr{specific} acoustic resistance $\mathrm{Re}(Z)$ becomes negative \rv{over a portion of the considered frequency range}, which implies amplification of the acoustic forcing by the bulk flow \rv{at the respective frequencies} \tr{\cite{bourquard_faure-beaulieu_noiray_2021}}. This occurs when the acoustic energy produced by the \tr{forcing of the convectively unstable shear layer} in the \tr{side branch aperture} exceeds the \tr{radiation} losses \tr{in the main branch}. \rv{Similar oscillating behavior of the acoustic impedance of the aperture was also observed in \cite{KOOIJMAN2008849}, where the effect of mean grazing flow on the acoustic response of a single rectangular slot in a wall to imposed sound was studied experimentally.} In \cite{bourquard_faure-beaulieu_noiray_2021}, the nonlinear saturation mechanism reported in \cite{boujo_bauerheim_noiray_2018} was also observed, manifesting itself in a flattening of the resistance curve for increasing $p_a$. This leads to a shrinking of the frequency range in which the resistance is negative until eventually \tr{it} becomes positive for the entire considered frequency \tr{range}. The authors further showed that using \tr{a second-order $\it{black}$-$\it{box}$ $\it{transfer}$ $\it{function}$ model}, a good fit over the frequency \tr{and forcing amplitude} ranges considered could be achieved. 

The present work is a continuation of this research, wherein the attempt is made to develop \tr{$\it{physics}$-$\it{based}$} models which can predict accurately the specific acoustic impedance $Z$ of the \tr{side branch opening} over a given frequency range for different grazing flow speeds $U$ and forcing amplitudes $p_a$. \tr{Here, we do not aim at predicting the aeroacoustics of the cavity \fc{opening} from the compressible Navier--Stokes equations directly. This is computationally expensive, even when LES of the turbulent flow, which constitutes already a significant reduction of the huge amount of degrees of freedom, is performed \cite{boujo_bauerheim_noiray_2018}}. Instead, \tr{our contribution falls in the category of simplified physics-based analytical models, classical examples of which were developed by \citet{howe_1998}}. To the knowledge of the authors, no 
\tr{$\it{physics}$-$\it{based}$} predictive models for the specific acoustic impedance of a harmonically forced aperture subject to a turbulent grazing flow of varying speed exist in literature \tr{that have been validated with experimental data}, especially ones that include the effect of \tr{large forcing amplitudes. This work therefore \fc{complements} and is located between \cite{boujo_bauerheim_noiray_2018} and \cite{bourquard_faure-beaulieu_noiray_2021} in the spectrum of modeling strategies for this cavity flow \fc{configuration}, the latter attacking the problem with a black-box modeling of the shear layer response to acoustic perturbations, and the former considering the Navier--Stokes equations to unravel the fundamental hydrodynamic mechanisms governing this response.} Because the aeroacoustic response of the \tr{side branch opening} is a key element of models describing \tr{the aeroacoustic instability responsible for} self-sustained cavity oscillations, the present work represents a significant contribution to various fields of research concerned with such instabilities.

The paper is structured as follows: in $\S$\ref{sec:: Section 2: Modelling}, we introduce the problem of modelling the acoustic response of the \tr{side branch aperture} by visualizing the hydrodynamic disturbance in the turbulent shear layer over the \tr{aperture} using PIV data for increasing \tr{forcing amplitudes} $p_a$. Then, we derive two models for the specific acoustic impedance of the \tr{opening}. The first model is based on Howe's classic formulation that models the shear layer as an infinitely thin vortex sheet that separates two fluid layers of constant but different \fc{mean} streamwise velocity, and the other on an assumed vertical velocity profile \fc{along} the \tr{side branch aperture}. In $\S$\ref{sec:: Section 3: Results}, we validate these two models by comparing their predictions of the specific acoustic impedance to the \tr{measurements, which were presented in \citet{bourquard_faure-beaulieu_noiray_2021}}. We then compare the \tr{representation of the hydrodynamic response and of the saturation effect by the two models} under increasing $p_a$. Finally, we discuss alternative models \fc{to the ones presented in this work}. In $\S$\ref{sec:: Section 4: Conclusions}, we summarize our conclusions.

\tr{\section{Modelling the acoustic impedance \label{sec:: Section 2: Modelling}}}
In this section, we derive two models for the acoustic impedance of the side branch \tr{aperture}. These models will from now on be referred to as model 1 \tr{and} 2. For model 1, we consider a right-handed coordinate system with origin located in the middle of the \tr{aperture}, where $x$ is the streamwise coordinate, $y$ is the vertical coordinate and is positive along the \tr{side branch}, and $z$ is the spanwise coordinate. The wind tunnel we consider has a cross-section area of $62\times62$ mm$^2$, with a side branch of width $W=30$ mm with the same spanwise extension \tr{$H=62$ mm} as the main channel. \rv{The experimental set-up is is presented in \cite{bourquard_faure-beaulieu_noiray_2021}, with a sketch of the overall set-up in Fig. 5(a) and with a picture of the side branch opening in Fig. 3.} We denote the cross-sectional area of the \tr{opening} by \tr{$A_o=HW$}. Throughout this paper, the ambient air density is $\rho_0=1.10$ kg/m$^3$ and the ambient speed of sound $c_0=350$ m/s. The total velocity field $\boldsymbol{v}(\boldsymbol{x},t)$ \fc{is decomposed} into \fc{its} time-averaged component $\bar{\boldsymbol{v}}(\boldsymbol{x})$, \fc{its} coherent fluctuations $\tilde{\boldsymbol{v}}(\boldsymbol{x},t)$, and \fc{its} turbulent fluctuations $\check{\boldsymbol{v}}(\boldsymbol{x},t)$: $\boldsymbol{v}(\boldsymbol{x},t)=\bar{\boldsymbol{v}}(\boldsymbol{x})+\tilde{\boldsymbol{v}}(\boldsymbol{x},t)+\check{\boldsymbol{v}}(\boldsymbol{x},t)=\langle \bar{\boldsymbol{v}}(\boldsymbol{x}) \rangle + \check{\boldsymbol{v}}(\boldsymbol{x},t)=\bar{\boldsymbol{v}}(\boldsymbol{x})+\boldsymbol{v}'(\boldsymbol{x},t)$, where \tr{$\langle\hspace{0.3mm}\cdot\hspace{0.3mm}\rangle$ denotes phase-averaging and} $\boldsymbol{v}'(\boldsymbol{x},t)$ are the zero mean fluctuations. \tr{Note that the notations used in this work differ somewhat from \cite{boujo_bauerheim_noiray_2018}.}

\begin{figure*}
\begin{psfrags}
\psfrag{a}{$\pi/2$}
\psfrag{b}{$\pi$}
\psfrag{c}{$3\pi/2$}
\psfrag{d}{$2\pi$}
\psfrag{g}{$300$ Pa}
\psfrag{h}{$y$ [mm]}
\psfrag{0}{0}
\psfrag{m}{-15}
\psfrag{j}{0}
\psfrag{x}{15}
\psfrag{o}{80}
\psfrag{p}{60}
\psfrag{q}{40}
\psfrag{r}{20}
\psfrag{k}{-10}
\psfrag{z}{$10$ Pa}
\psfrag{u}{$50$ Pa}
\psfrag{i}{$x$ [mm]}
\psfrag{f}{$\langle v_x \rangle $ [m/s]}
  \centerline{\includegraphics{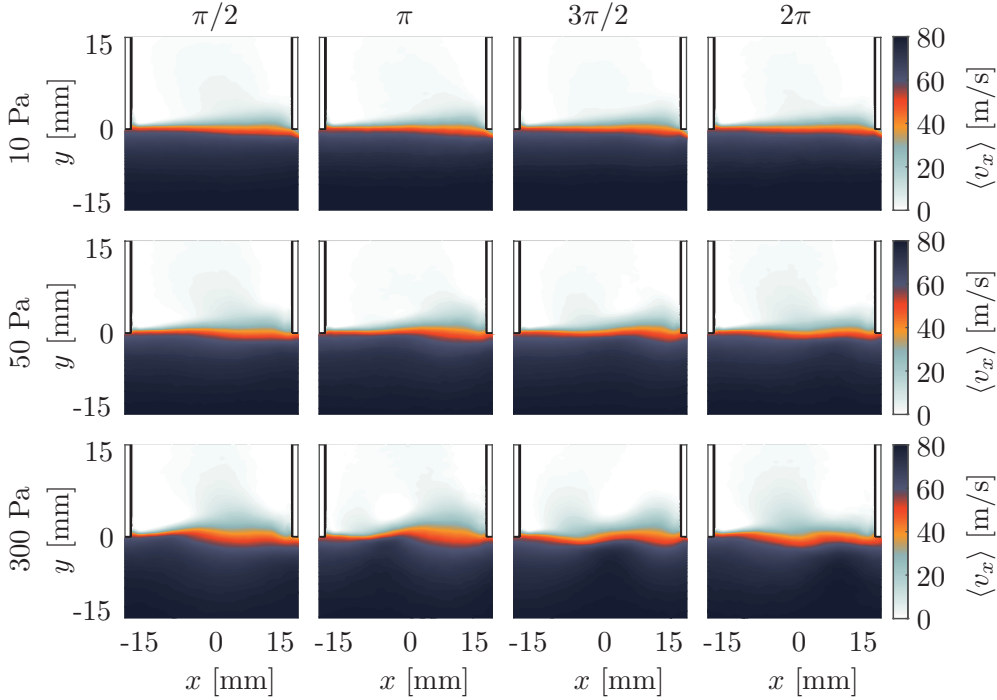}}
    \end{psfrags}

  \caption{PIV data obtained from the center plane of the side branch aperture visualizing the spatiotemporal evolution of the hydrodynamic disturbance in the shear layer for a mean bulk flow velocity $U=74.1$ m/s under acoustic forcing with frequency $f=1500$ Hz and acoustic pressure forcing amplitudes $p_a$ of $10$ Pa (top row), $50$ Pa (middle row) and $300$ Pa (bottom row), respectively. Shown is the phase-averaged streamwise velocity $\langle v_x(\boldsymbol{x},t) \rangle=\bar{v}_x(\boldsymbol{x})+\tilde{v}_x(\boldsymbol{x},t)$ at four equally spaced time instants over a full acoustic forcing cycle. The phase $\omega t$ of the acoustic forcing is displayed above the frames in the top row.}
\label{fig:Figure 1}
\end{figure*}

\begin{figure*}
\begin{psfrags}
\psfrag{a}{$\pi/2$}
\psfrag{b}{$\pi$}
\psfrag{c}{$3\pi/2$}
\psfrag{d}{$2\pi$}
\psfrag{g}{$300$ Pa}
\psfrag{h}{$y$ [mm]}
\psfrag{0}{0}
\psfrag{m}{-15}
\psfrag{j}{0}
\psfrag{x}{15}
\psfrag{o}{10}
\psfrag{p}{60}
\psfrag{q}{\hspace{.5mm}0}
\psfrag{r}{-10}
\psfrag{k}{-10}
\psfrag{z}{$10$ Pa}
\psfrag{u}{$50$ Pa}
\psfrag{i}{$x$ [mm]}
\psfrag{f}{\hspace{3mm}$ \tilde{\omega}_z  $ [s$^{-1}$]}
  \centerline{\includegraphics{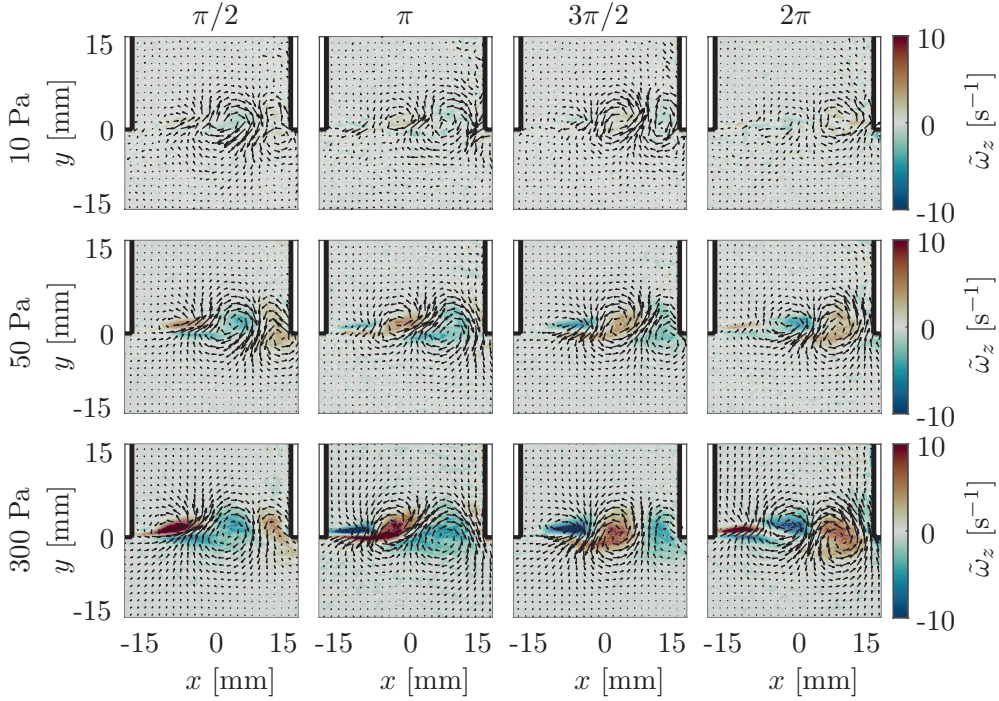}}
    \end{psfrags}

  \caption{\tr{PIV data obtained from the center plane of the side branch aperture visualizing the spatiotemporal evolution of the hydrodynamic disturbance in the shear layer for a mean bulk flow velocity $U=74.1$ m/s under acoustic forcing with frequency $f=1500$ Hz and acoustic pressure forcing amplitudes $p_a$ of $10$ Pa (top row), $50$ Pa (middle row) and $300$ Pa (bottom row), respectively. Shown is the vector field of the coherent velocity fluctuations $\tilde{\boldsymbol{v}}(\boldsymbol{x},t)$, superimposed on the coherent vorticity fluctuations $\tilde{\omega}_z(\boldsymbol{x},t)$ at four equally spaced time instants over a full acoustic forcing cycle. The phase $\omega t$ of the acoustic forcing is displayed above the frames in the top row.}}
\label{fig:Figure 2}
\end{figure*}

PIV data obtained from the center plane of the side branch aperture is used in \tr{\fc{Figs.} \ref{fig:Figure 1} and \ref{fig:Figure 2}} to visualize the spatiotemporal evolution of the hydrodynamic disturbance in the shear layer for a mean bulk flow velocity $U=74.1$ m/s under \tr{acoustic forcing} with frequency $f=1500$ Hz and acoustic \tr{pressure forcing} amplitudes $p_a$ of $10$ Pa, $50$ Pa and $300$ Pa, respectively. \tr{\fc{Figure} \ref{fig:Figure 1} shows the phase-averaged streamwise velocity $\langle v_x(\boldsymbol{x},t) \rangle=\bar{v}_x(\boldsymbol{x})+\tilde{v}_x(\boldsymbol{x},t)$ and Fig. \ref{fig:Figure 2} the vector field $\tilde{\boldsymbol{v}}(\boldsymbol{x},t)$, superimposed on the coherent vorticity fluctuations $\tilde{\omega}_z(\boldsymbol{x},t)=\partial \tilde{v}_y(\boldsymbol{x},t)/\partial x-\partial \tilde{v}_x(\boldsymbol{x},t)/\partial y$, respectively,} at four equally spaced time instants over a full acoustic forcing cycle. The phase $\omega t$, where $\omega=2\pi f$, of the acoustic forcing is displayed above the frames in the top row. In Fig. 1, a disturbance in the coherent streamwise velocity is visible that grows more pronounced with increasing \tr{forcing} amplitude. \tr{In Fig. 2, we observe \fc{shedding of coherent vorticity fluctuations} that changes from a spurious to a clearly discernible pattern with increasing $p_a$. However, even at $p_a=300$ Pa, no roll-up of discrete vortices takes place.} \fc{It is worth mentioning that at this forcing amplitude, the acoustic velocity is about $0.8$ m/s at the aperture, and the associated vertical acoustic displacement is about $80$ $\mu$m.} \tr{These features of the coherent velocity and vorticity fields illustrate the thickening of the mean shear layer for large forcing amplitudes which reduces the shear driven amplification of the disturbances \fc{from} the upstream edge of the \tr{side branch opening}.}

\tr{Model 1 is based on Howe's vortex sheet theory \cite{Howe1996} and model 2 on an assumed coherent} vertical velocity \tr{profile $\tilde{v}_y$ in the side branch aperture.} \tr{It is not a straightforward task to compute the acoustic impedance in the presence of a mean flow, because, as \citet{hirschberg2007introduction} states, "when the reference fluid is not uniform or there is a mean flow, there
is some arbitrariness in the definition of the acoustical field and of the corresponding acoustical energy".} \tr{In our approach, we compute $Z$ from} the Rayleigh conductivity $K_R$ \cite{Howe1996}. When a harmonic pressure load $p_a e^{-i \omega t}$\fc{, with $p_a$ a real positive constant,} is applied to the side branch opening and causes a coherent volume flux $Q(t)$ in positive $y$-direction, \rv{$K_R$ is defined, see eq. (5.3.1) in \cite{howe_1998}, as 
\begin{equation}
    K_R=-\frac{i \omega \rho_0 Q}{p_a}, \label{Rayleigh conductivity}
\end{equation}}
where $i$ is the imaginary unit and $Q$ is the coherent volume flux \tr{through the aperture \fc{and $K_R$ is a length}.} In Howe's theory, see eq. (2.4) in \cite{Howe1996}, \rv{$Q=-i\omega\int_{A_o} \zeta dS$}, where $\zeta$ is the (complex) vortex sheet displacement and the \rv{integral is taken over the aperture area $A_o$.} Both $Q$ and the pressure load vary like $e^{-i\omega t}$ over time, which makes this factor cancel out in eq. (\ref{Rayleigh conductivity}). In contrast, for model 2, we derive $K_R$ from a real pressure load $p_a \cos \omega t$ and a real coherent vertical velocity \fc{profile in the aperture $\tilde{v}_y(x,t)$, which we denote by $v_{y,c}(x,t)$.} \rv{To make an analogy to Howe's theory and eq. (\ref{Rayleigh conductivity}), we define, for model 2,
\begin{equation}
    Q=s \int_{A_o} \hat{v}_{y,c}\hspace{1mm}dS, \label{Volume flux analogy}
\end{equation} 
where $\hat{f}(s)=\mathcal{L}[f(t)](s)$ denotes the Laplace transform \rv{\cite{debnath2014integral}} of a function $f(t)$, $s$ is the Laplace variable, and we set $s\equiv i\omega$ to obtain the frequency response. Note that $Q$ defined in eq. \eqref{Volume flux analogy} has the units of a volume flux.} From $K_R$, we compute the specific acoustic impedance $Z$ as follows:
\begin{equation}
    \rv{Z=-\frac{i \omega A_o}{c_0 K_R}=\frac{A_o p_a}{\rho_0 c_0 Q}.} \label{Acoustic impedance}
\end{equation}
\rv{With this formulation, it is clear that for $s=i\omega$, $\omega\in\mathbb{R}$, a positive imaginary part of the Rayleigh conductivity $K_R$ implies a reflection coefficient $R=(1-Z)/(1+Z)$ with magnitude $|R|>1$, which indicates amplification of the sound field by the mean flow \cite{Howe1996,Howe97,howe_1998}.} Note that in the absence of a mean flow, when $Q=A_o v_a$, where $v_a$ is the acoustic velocity amplitude, eq. (\ref{Acoustic impedance}) coincides with Hirschberg's definition of the acoustic impedance, given by eq. (88) in \cite{hirschberg2007introduction}.\\

\subsection{Model 1\label{sec::Modification of Linear Model}}

\begin{figure*}
\begin{psfrags}
\psfrag{H}{\it{H}}
\psfrag{U}{$U$}
\psfrag{Q}{$Q=s\left(\int_{A_1} \hat{v}_{y,c}\hspace{1mm}dS+\int_{A_2} \hat{v}_{y,c}\hspace{1mm}dS+\int_{A_3} \hat{v}_{y,c}\hspace{1mm}dS\right)$}
\psfrag{W}{\it{W}}
\psfrag{v1}{$v_{y,1}$}
\psfrag{w}{$U_+$}
\psfrag{s}{$U_-$}
\psfrag{k}{\,\,\,$v_{y,c}$}
\psfrag{z}{$\zeta_R$}
\psfrag{v}{$\hspace{2mm}A_1$}
\psfrag{i}{$\hspace{2mm}A_2$}
\psfrag{l}{$\hspace{2mm}A_3$}
\psfrag{u}{$u_c$}
\psfrag{g}{$\xi$}
\psfrag{y}{$y$}
\psfrag{x}{$x$}
\psfrag{a}{(a)}
\psfrag{b}{(b)}
\psfrag{p}{\,\,\,$p_a\cos{\omega t}$}
  \centerline{\includegraphics[width=0.4\textwidth]{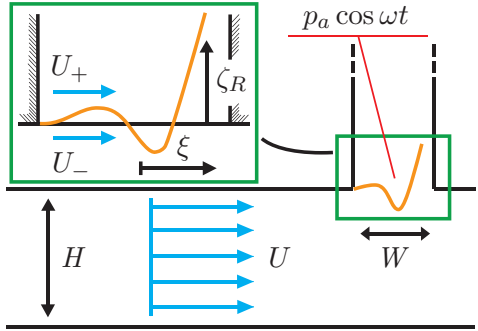}}
  \end{psfrags}
  \caption{Sketch of the experimental configuration. The input parameters of model 1 are shown: the mean bulk flow velocity $U$, the \tr{side branch} width $W$, the channel height $H$ and the acoustic pressure forcing $p_a \cos{\omega t}$ that is applied \tr{above the side branch opening}. In the cutout, the real part of the vortex sheet displacement $\zeta_R(\xi,t)=\mathrm{Re}{(\zeta(\xi) e^{-i\omega t})}$\fc{, the mean streamwise velocities just above and below the vortex sheet, $U_+$ and $U_-$, respectively,} and the scaled streamwise coordinate $\xi=2x/W$ are shown. }
\label{fig:Figure 3}
\end{figure*}
In this section, we derive model 1, which is based on Howe's classic formulation that describes the shear layer as a thin vortex sheet which separates two regions of constant but different mean streamwise \tr{velocity}. For more details regarding the theory behind this model, the reader is referred to \tr{\cite{howe1979theory,howe_1981,Howe1996,Howe97} and chapters 5 and 6 of \cite{howe_1998}}. The input parameters of model 1 and the experimental setting are shown in Fig. \fc{\ref{fig:Figure 3}}: the mean bulk flow velocity $U$, the \tr{side branch} width $W$, the channel height $H$ and the acoustic pressure \tr{load} \tr{$\mathrm{Re}(p_a e^{-i \omega t})=p_a \cos{\omega t}$}\fc{,} which is applied \tr{across} the \tr{side branch opening}. This forcing, as indicated in Fig. \tr{\ref{fig:Figure 3}}, causes a time-harmonic displacement \tr{$\zeta_R(\xi,t)=\mathrm{Re}{(\zeta(\xi) \rv{e^{-i\omega t}})}$} of the vortex sheet, where $\xi=2x/W$ is the scaled streamwise coordinate. In the cutout, \tr{$\zeta_R(\xi,t)$}\fc{, the mean streamwise velocities just above and below the vortex sheet, $U_+$ and $U_-$, respectively,} and $\xi$ are shown. Following \cite{howe_1998}, the dependence of the velocity field in the aperture on the spanwise variable is neglected, which is why it is sufficient to only consider a cross-section as we do in this work. Fig. \fc{\ref{fig:Figure 3}} also illustrates the the Kutta condition \cite{howe_1981}, which states that the vortex sheet is tangential to the main \tr{duct} at the upstream edge of the \tr{side branch opening}. The derivation of Howe's model\tr{, which is taken here as a basis for model 1,} is given in appendix \ref{Appendix A}. This derivation\rv{, which starts from the linearized unsteady Bernoulli equation}, leads to the following equation:
\begin{equation}
    \int_{-1}^1 \zeta'(\mu) \left[\ln{|\xi-\mu|}+L_-(\xi,\mu)+K(\xi,\mu)\right] d\mu+\lambda_1 e^{i \sigma_1 \xi}+\lambda_2 e^{i \sigma_2 \xi}=1, \label{Master equation for zeta (Section 2)}
\end{equation}
where \tr{$\zeta'=\zeta\rho_0\omega^2W/\pi p_a$}, $\mu$ is an integration variable corresponding to $\xi$, $\sigma_{1,2}=\sigma(1\pm i)/(1\pm i\alpha)$, \tr{$\sigma=\omega W/2 U_-$} is the Kelvin-Helmholtz wavenumber, \tr{$\alpha=U_+/U_-$} and \tr{the functions $L_-$ and $K$ are defined in appendix \ref{Appendix A}}. Equation (\ref{Master equation for zeta (Section 2)}) is \tr{an} integral equation which \tr{is here} solved numerically for $\zeta'(\xi)$ and the constants $\lambda_{1,2}$ subject to the Kutta condition 
\begin{equation}
    \zeta'(-1)=\frac{\partial \zeta'}{\partial \xi}(-1)=0. \label{Kutta condition, Section 2}
\end{equation} 
\tr{The method used for the solution of eq. (\ref{Master equation for zeta (Section 2)}) is detailed in appendix \ref{Appendix B}.} From this solution, we then obtain the Rayleigh conductivity according to the following formula\rv{, given by eq. (6.1.10) in \cite{howe_1998}}:
\tr{\begin{equation}
    \rv{K_R=-\frac{\pi H}{2}\int_{-1}^1 \zeta'(\mu) d\mu.} \label{Rayleigh conductivity, original Howe model}
\end{equation}}\tr{A few more parameters are now added to the model such that, after calibration, prediction of the specific acoustic impedance of the opening can be made over broad ranges of the frequency, bulk flow velocity $U$ and acoustic pressure forcing amplitude $p_a$.} First, we adjust $U_+$ and $U_-$ by introducing the following relations: 
\tr{\begin{eqnarray}
    U_+&&= \left[\alpha_0+\alpha_1\left(1-\frac{p_a}{p_{a,0}}\right)\right] U_- ,\label{Modification 1: U_-}\\
    U_-&&=\beta U. \label{Modification 2: sigma}
\end{eqnarray}}
The modification (\ref{Modification 1: U_-}) means that there is a small mean streamwise velocity above the shear layer which is caused by recirculation cells that form in the \tr{side branch}. For increasing acoustic pressure forcing amplitude $p_a$\tr{, starting from a small amplitude $p_{a,0}$ in the linear regime,} the shear layer thickens \tr{\cite{boujo_bauerheim_noiray_2018} and} we expect the mean flow, and especially the ratio $U_+/U_-$, to change. This \tr{amplitude-dependent effect} is \tr{modelled with} the parameter $\alpha_1$. Equation (\ref{Modification 2: sigma}) implies that the mean flow velocity just below the vortex sheet $U_-$\tr{, which depends on the velocity profile of the boundary layer in the main duct upstream of the side branch,} is set to \tr{$\beta U$}. Additionally, we introduce a complex offset for the \fc{specific} acoustic impedance
\begin{equation}
    Z\rightarrow Z+\tr{\gamma_1}+i \omega \delta. \label{Modification 3: Offsets}
\end{equation}
The two constants \tr{$\gamma_1$} and $\delta$ account for \tr{corrections of} radiation losses and \tr{inertial} effects \tr{at the side branch} opening.

We have now derived model 1, which is defined by the solution $\zeta(\xi)$ of (\ref{Master equation for zeta (Section 2)}), where $\xi \in \left[-1,1\right]$, which satisfies the Kutta condition (\ref{Kutta condition, Section 2}). Once we have computed $\zeta$, we use eq. (\ref{Rayleigh conductivity, original Howe model}) to obtain $K_R$. Substituting this value into (\ref{Acoustic impedance}), and adding the offsets \tr{$\gamma_1$} and $i\omega\delta$ defined in (\ref{Modification 3: Offsets}) yield for each forcing frequency $f$ a value of the specific acoustic impedance of the side branch opening $Z$ that can be compared to data obtained from the measurements:
\begin{equation}
    Z=\frac{i \omega \tr{A_o}}{c_0 K_R}+\tr{\gamma_1}+i \omega \delta. \label{Acoustic impedance with offsets}
\end{equation}
In total, \rv{model 1 contains 5 empirical parameters}: 4 of these parameters, $\alpha_0$, $\beta$, $\gamma_1$ and $\delta$ are included to achieve good predictions for different bulk flow velocities $U$, and are determined from a fit to the \tr{measured} real and imaginary impedance curves $Z(\omega)$ for the grazing flow speed $U=74.1$ m/s at a small forcing amplitude $p_{a,0}=20$ Pa. The last parameter $\alpha_1$ represents the \fc{alteration} of the mean flow by the acoustic forcing, and is determined from a similar fit, for the same $U$, at a high forcing pressure amplitude, in our case at $p_a=800$ Pa. \tr{It is worth noting that the actual 3D shear layer dynamics are not explicitly described in this model, which considers an idealized 2D vortex sheet, but that these 3D effects will affect the value of the calibrated parameters. This \rv{is} further discussed in $\S$\ref{sec:: Section 4: Discussion}}.

\subsection{Model 2 \label{Subsection model b}}
\begin{figure*}
\begin{psfrags}
\psfrag{H}{\it{H}}
\psfrag{U}{$U$}
\psfrag{Q}{$Q=s\left(\int_{A_1} \hat{v}_{y,c}\hspace{1mm}dS+\int_{A_2} \hat{v}_{y,c}\hspace{1mm}dS+\int_{A_3} \hat{v}_{y,c}\hspace{1mm}dS\right)$}
\psfrag{W}{\it{W}}
\psfrag{v1}{$v_{y,1}$}
\psfrag{w}{$U_+$}
\psfrag{s}{$U_-$}
\psfrag{k}{\,\,\,$v_{y,c}$}
\psfrag{z}{$\zeta_R$}
\psfrag{v}{$\hspace{2mm}A_1$}
\psfrag{i}{$\hspace{2mm}A_2$}
\psfrag{l}{$\hspace{2mm}A_3$}
\psfrag{u}{$u_c$}
\psfrag{g}{$\xi$}
\psfrag{y}{$y$}
\psfrag{x}{$x$}
\psfrag{a}{(a)}
\psfrag{b}{(b)}
\psfrag{p}{\,\,\,$p_a\cos{\omega t}$}
  \centerline{\includegraphics[width=0.4\textwidth]{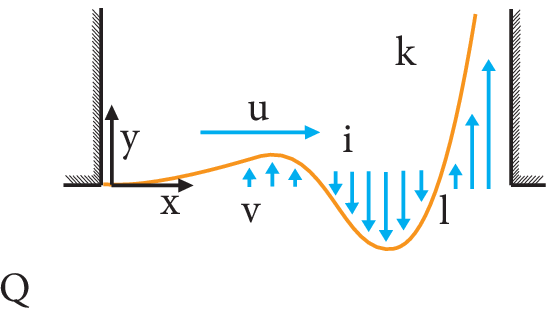}}
  \end{psfrags}
  \caption{\rv{Sketch of model 2.} The coordinate system used for model 2 is shown, as well the \tr{convective speed of the perturbations} $u_c$ and a typical distribution of \tr{the coherent vertical velocity field in the aperture \fc{$v_{y,c}(x,t)$}} at a given time instant $t$. The contributions of the different regions in the aperture to the coherent volume flux $Q$ defined in eq. (\ref{Volume flux analogy}) are also indicated. \fc{Note that the origin is not the same for \rv{Figs. \ref{fig:Figure 3} and \ref{fig:Figure 4}.}}}
\label{fig:Figure 4}
\end{figure*}
In this section, we derive model 2, an analytical model of the acoustic impedance of the \tr{side branch} opening. Our approach is inspired by \citet{takahashi2016theoretical}, who use an assumed vertical velocity profile to qualitatively estimate the acoustic energy produced by an oscillating jet. \tr{Here, we assume a coherent vertical velocity profile in the shear layer that develops at the side branch opening under the turbulent grazing flow.} In this model, the origin of the coordinate system is placed at the upstream corner. A harmonic acoustic forcing $p_a \cos{\omega t}$ is applied \tr{across} the \tr{side branch} opening. We assume that this acoustic pressure field and the turbulent grazing flow lead to the following \tr{coherent} displacement field \tr{$\tilde{y}(x,t)$ in the aperture}:
\begin{equation}
    \tr{\tilde{y}(x,t)=\frac{v_a }{\omega} g(x)\sin{\omega (t-x/u_c)}} \label{Particle displacement in aperture},
\end{equation}
where $v_a=p_a/\rho_0 c_0$, and \tr{$u_c=\kappa U$} is the grazing flow speed in the aperture, and $\kappa$ is \tr{a parameter} that describes the ratio of \tr{$u_c$ to $U$}. For the function $g(x)$, we choose a truncated polynomial of order $N$:
\begin{equation}
    g(x)=H(x)\left[(a_0 + a_1 x)\left(1-\frac{p_a}{p_{a,0}}\right)+\sum^N_{k=2} \frac{a_k}{k!}x^k\right], \label{g function 1}
\end{equation}
where $H(x)$ is the Heaviside function. We chose $N=5$ in this work. This large number of parameters in the function $g$ is needed to \rv{achieve} robust predictions of the specific acoustic impedance over wide ranges of frequency, bulk velocity and acoustic pressure forcing amplitude. Model 2 is sketched in Fig. \tr{\ref{fig:Figure 4}}. In the figure, the coordinate system used for model 2 is shown, as well as $u_c$ and a typical distribution of $v_{y,c}(x,t)$ at a given time instant $t$. The same figure also indicates the contributions of the different regions in the aperture to the coherent volume flux $Q$ defined in eq. (\ref{Volume flux analogy}). 

For the streamwise component of the velocity of a fluid particle in the aperture with initial condition $(x_0,0)$ at time $t=0$, we have, approximately, $v_x(x,y=0,t) =u_c$, which gives, for the particle motion, \tr{$x(t)=u_c t + x_0$}. Substituting this result into eq. (\ref{Particle displacement in aperture}) \tr{with $x\equiv x_p(t)$,} and taking its partial time derivative, we obtain the $y$-component of the coherent fluid particle velocity in the \tr{side branch} aperture in Lagrangian coordinates:
\begin{eqnarray}
    v_{y,c}(x_p(t),t)=&&\frac{\partial}{\partial t} \fc{\tilde{y}}(x_p(t),t)\\
    =&&\frac{\fc{v_a} }{\omega}\frac{\partial}{\partial t} \left( g(x_p)\sin{\omega (t-x_p/u_c)}\right)\\
    =&&-\frac{\fc{v_a} u_c }{ \omega}g'( x_p)\sin{\omega x_0/u_c}.
    \label{Lagrangian velocity field}
\end{eqnarray}
Taking the Laplace transform of this expression \tr{and setting $s\equiv i \omega$, where $s$ is the Laplace variable,} yields the vertical velocity in the aperture  \tr{plane} in the frequency domain, \tr{$\hat{v}_{y,c}(x_0,s)$}:
\begin{eqnarray}
 \hat{v}_{y,c}(x_0,s)=\mathcal{L}[v_{y,c}](x_0,s)=&&- \frac{\fc{v_a} u_c }{ \omega} \mathcal{L}[g'(u_c t+x_0)](s)\sin{\omega x_0/u_c}\\
 =&&- \frac{\fc{v_a}  }{ \omega} \mathcal{L}[g'(t+x_0)](s/u_c)\sin{\omega x_0/u_c}\\
 =&& -\frac{\fc{v_a} }{ \omega} \mathcal{L}[g'(t)]\left(s/u_c\right)e^{ x_0 s/u_c}\sin{\omega x_0/u_c}. \label{Lagrangian aperture velocity field in frequency domain}
\end{eqnarray}
The \tr{volume flux $Q$ we defined in eq. (\ref{Volume flux analogy}) is}
\begin{equation}
    \tr{Q(s)=s\int_{A_o} \hat{v}_{y,c}(x,s) dS,}
\end{equation}
where the integral is taken over the aperture area \tr{$A_o:\{0\leq x\leq W;0\leq z\leq H\}$}, which is fixed in space at all times. Since we know the velocity field in the aperture in Lagrangian coordinates, we write this integral in the initial configuration. Hence, we have
\tr{\begin{eqnarray}
Q(s)=&&s\int_{A_o} \hat{v}_{y,c}(x_0,s)d x_0 d z_0\\
=&&sH\int_0^W \hat{v}_{y,c}(x_0,s) d x_0\\
 =&&-\frac{s H v_a }{ \omega}\mathcal{L}[g'(t)](s/u_c)\int_0^W  e^{ s x_0/u_c}\sin{\omega x_0/u_c} d x_0\\
 =&&-\frac{sH v_a  }{ 2 i \omega}\mathcal{L}[g'(t)]\left(s/u_c\right)\int_0^W  \left(e^\frac{2 s x_0 }{u_c}-1\right) d x_0\\
 =&&-\frac{s H v_a }{ 2 i \omega}\mathcal{L}[g'(t)]\left(s/u_c\right)\left(\frac{u_c}{2 s}\left[e^\frac{2sW}{u_c}-1\right]-W\right).
\end{eqnarray}}Substituting this result into the definition of the \tr{specific} acoustic impedance (\ref{Acoustic impedance}), and adding to the result a correction term given by the real offset \tr{$\gamma_2$}, which accounts for radiation losses and 3D effects, yields
\begin{eqnarray}
    Z(s)=\frac{4sW }{\left(2sW+u_c\left[1-e^\frac{2sW}{u_c}\right]\right)}\frac{1}{\mathcal{L}[g'(t)]\left(s/u_c\right)}+\gamma_2. \label{Acoustic impedance, model 2}
\end{eqnarray}
Note that for \tr{$g$} given by \tr{eq.} (\ref{g function 1}) with $N=5$, we get the following expression for $\mathcal{L}[g'(t)](s/ u_c)$:
\begin{equation}
    \mathcal{L}[g'(t)](s/u_c)=\left[a_0+a_1\left(\frac{u_c}{s}\right)\right]\left(1-\frac{p_a}{p_{a,0}}\right)+\sum^5_{k=2} a_k\left( \frac{u_c}{s} \right)^k.
\end{equation}
In total, model 2 contains 8 empirical parameters. The coefficients $a_k, \hspace{1mm} k\geq 2$ and $\kappa$ are calibrated from a fit to the real and imaginary impedance curves $Z(\omega)$ for the mean bulk flow velocity $U=74.1$ m/s at a small \tr{acoustic pressure forcing} amplitude $p_{a,0}=20$ Pa. The coefficients $a_0$ and $a_1$, which describe the dependence of the specific impedance $Z$ on the \tr{forcing amplitude} $p_a$, are determined from a similar fit, for the same $U$, at a \tr{large forcing} amplitude \rv{$p_{a}=800$} Pa.

\section{Results\label{sec:: Section 3: Results}}
In this section, we present and discuss the results obtained from the models derived in $\S$\ref{sec:: Section 2: Modelling}. Two types of data are analyzed in this section: On one hand, models 1 and 2 predict values of the \tr{specific} acoustic impedance $Z$ of the \tr{side branch aperture}. These values are compared to the experimentally measured values of $Z$ from \cite{bourquard_faure-beaulieu_noiray_2021}. On the other hand, models 1 and 2 also describe the hydrodynamic response of the shear layer, represented by the velocity induced by the vortex sheet displacement, \tr{$\mathrm{Re}(-i\omega \zeta(x)e^{-i \omega t})$} and the velocity field $v_{y,c}(x,t)$, respectively. We compare these two velocity fields in the aperture for increasing forcing pressure amplitudes $p_a$ to investigate \tr{the models' predictions from a hydrodynamic perspective}. PIV data was not available in the relevant frequency range for comparison. \rv{This is because the experiments were performed at an earlier time, before the present study was initiated.} \tr{In table \ref{tab:table1}, we show the nondimensionalized values of the empirical parameters in model 1 and 2 so that the reader can reproduce the results presented in this section. The parameters were calibrated to the measured values of the specific acoustic impedance from \cite{bourquard_faure-beaulieu_noiray_2021}, as detailed in $\S$\ref{sec:: Section 2: Modelling}.}
\begin{table}[b]\tr{
\caption{\label{tab:table1}%
Nondimensionalized values of the empirical parameters in model 1 and 2. The values were calibrated to the measured values of the specific acoustic impedance from \cite{bourquard_faure-beaulieu_noiray_2021}, as detailed in $\S$\ref{sec:: Section 2: Modelling}.}
\begin{ruledtabular}
\begin{tabular}{cccc}
\textrm{Parameter (Model 1)}&
\textrm{Value}&
\multicolumn{1}{c}{\textrm{Parameter (Model 2)}}&
\textrm{Value}\\
\colrule
$\alpha_0$ & $0.149$ & $a_0$ & $ -7.54$\\
$\alpha_1$ & $4.63 \times 10^{-3}$ & $a_1 W$ & $-4.24\times 10^{-2}$\\
$\beta$ & $0.504$ & $a_2 W^2$ & $-205$\\
$\gamma_1$ & $0.337$ & $a_3 W^3$ & $ 183$\\
$\delta c_0/W$ & $-0.663$ & $a_4 W^4$ & $-3.30 \times 10^{3}$\\
{} & {} & $a_5 W^5$ & $6.28 \times 10^{3}$\\
{} & {} & $\kappa$ & $0.705$\\
{} & {} & $\gamma_2$ & $0.122$\\
\end{tabular}
\end{ruledtabular}}
\end{table}
\subsection{Prediction of the specific acoustic impedance} 
In Fig. \ref{fig:Figure 3}, we compare the predictions of $Z$ obtained with models 1 and 2 to the experimentally measured values for different bulk flow velocities $U$ at $p_a=20$ Pa. The parameters for models 1 and 2 were calibrated, as described in the previous section, to experimental data of the specific acoustic impedance $Z$ from \cite{bourquard_faure-beaulieu_noiray_2021} for the grazing flow speed $U=74.1$ m/s at low and high acoustic pressure forcing amplitudes $p_{a,0}=20$ Pa and $p_a=800$ Pa, respectively. The curves for which \tr{the calibration} of the empirical parameters in the models was performed are indicated by an increased line thickness compared to the other cases. Note that the frequency \tr{ranges} shown for the \rv{two lowest} bulk flow velocities are different than for the other cases. In general, we see that there is good agreement between models 1 and 2 and the experiments. Both models deviate from the measured values as we move away from the case $U=74.1$, for which the calibration was performed. Model 1 is unable to predict well $\mathrm{Im}{(Z)}$ at high frequencies. In contrast to this, the necessary condition for aeroacoustic instability $\mathrm{Re}{(Z)}<0$ is captured well by model 1 even for the smallest $U$, while model 2 does not describe correctly the shrinking of the domain where $\mathrm{Re}{(Z)}<0$ at lower $U$, which is visible at $U=56.5$ m/s and $U=60.0$ m/s.
\begin{figure*}
\begin{psfrags}
\psfrag{d}{$U=56.5$ m/s}
\psfrag{e}{$U=60.0$ m/s}
\psfrag{f}{$U=63.5$ m/s}
\psfrag{g}{$U=67.1$ m/s}
\psfrag{h}{$U=70.6$ m/s}
\psfrag{i}{$U=74.1$ m/s}
\psfrag{a}{$\mathrm{Re}{(Z)}$}
\psfrag{b}{$\mathrm{Im}{(Z)}$}
\psfrag{K}{6 \hspace{5.3mm}7}
\psfrag{E}{8 \hspace{4mm} 9}
\psfrag{G}{10}
\psfrag{H}{11}
\psfrag{I}{12}
\psfrag{J}{13}
\psfrag{A}{1}
\psfrag{B}{0.5}
\psfrag{C}{0}
\psfrag{Z}{$\times 10^2$}
\psfrag{c}{Frequency [Hz]}
  \centerline{\includegraphics{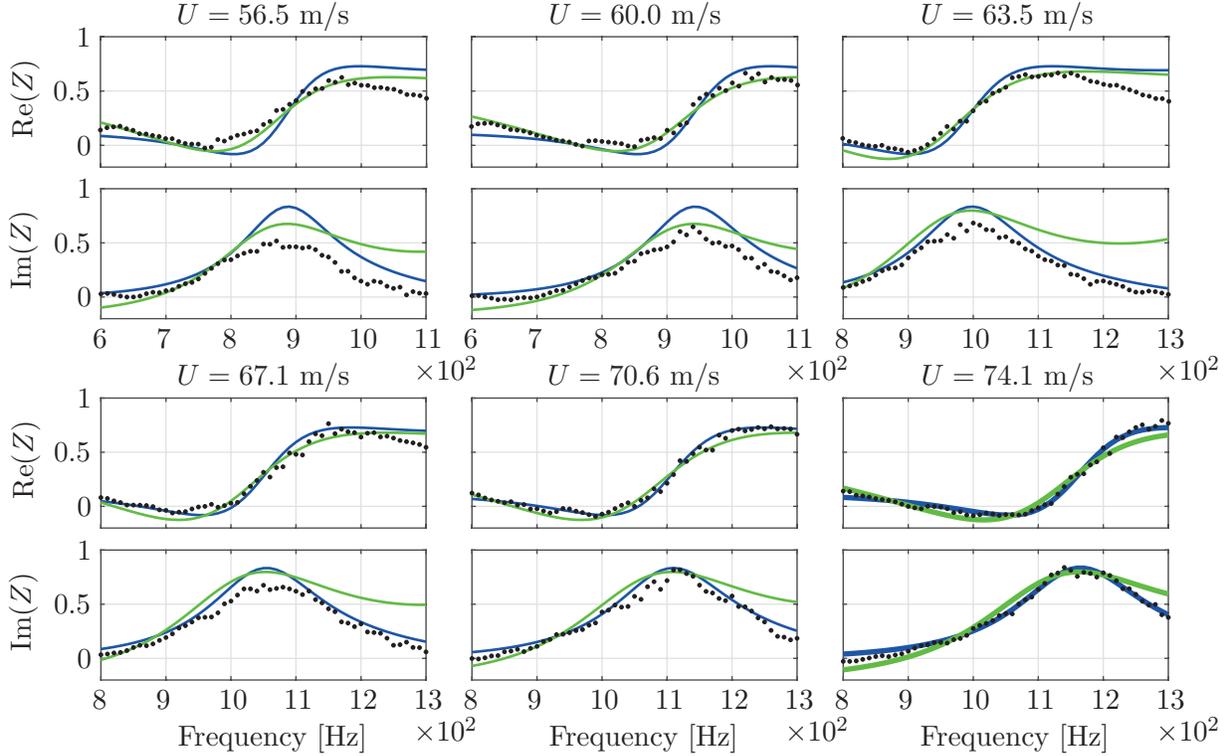}}
\end{psfrags}
  \caption{Comparison of the specific acoustic impedance \fc{$Z$} of the side branch opening, computed from model 1 (solid green line) and model 2 (solid blue line) with the experimentally measured values of the same from \cite{bourquard_faure-beaulieu_noiray_2021} (black dots). Shown are the real (first and third row) and imaginary parts (second and fourth row) of $Z$ for different bulk flow velocities $U$ for the acoustic pressure forcing amplitude $p_a=20$ Pa. The model parameters were calibrated to the data for the case $U=74.1$ m/s. Note that the frequency range considered for the cases $U=56.5$ m/s and $U=60.0$ m/s is different than for the other cases. The curves for which \tr{the calibration} of the empirical parameters in the models was performed are indicated by an increased line thickness compared to the other cases.}
\label{fig:Figure 5}
\end{figure*}
In \fc{Fig. \ref{fig:Figure 6}}, we compare the values of $Z$ predicted by model 1 and 2 to the measured values of the same for different acoustic pressure forcing amplitudes $p_a$ \tr{at the bulk flow velocity $U=74.1$ m/s}. The curves for which \tr{the calibration} of the empirical parameters in the models was performed are indicated by an increased line thickness compared to the other cases. We note that model 1 outperforms model 2 at the highest value of $p_a$, where the \tr{calibration was} performed. In general, however, both models capture well the nonlinear saturation effect, represented by a flattening of the curves $\mathrm{Re}{(Z)}(\omega)$ and $\mathrm{Im}{(Z)}(\omega)$ with increasing $p_a$. Especially, both models describe well the shrinking and eventual disappearance of the frequency \tr{range} for which the necessary conditon for instability $\mathrm{Re}{(Z)}<0$ is satisfied. Hence models 1 and 2 are able, after calibration, to quantitatively predict the saturation effect over nearly three orders of magnitude of the \tr{forcing amplitude}. 
\begin{figure*}
\begin{psfrags}
\psfrag{d}{$p_a=20$ Pa}
\psfrag{e}{$p_a=50$ Pa}
\psfrag{f}{$p_a=100$ Pa}
\psfrag{g}{$p_a=200$ Pa}
\psfrag{h}{$p_a=400$ Pa}
\psfrag{i}{$p_a=800$ Pa}
\psfrag{a}{$\mathrm{Re}{(Z)}$}
\psfrag{b}{$\mathrm{Im}{(Z)}$}
\psfrag{K}{6 \hspace{5.3mm}7}
\psfrag{E}{8 \hspace{4mm} 9}
\psfrag{G}{10}
\psfrag{H}{11}
\psfrag{I}{12}
\psfrag{J}{13}
\psfrag{A}{1}
\psfrag{B}{0.5}
\psfrag{C}{0}
\psfrag{Z}{$\times 10^2$}
\psfrag{c}{Frequency [Hz]}
  \centerline{\includegraphics{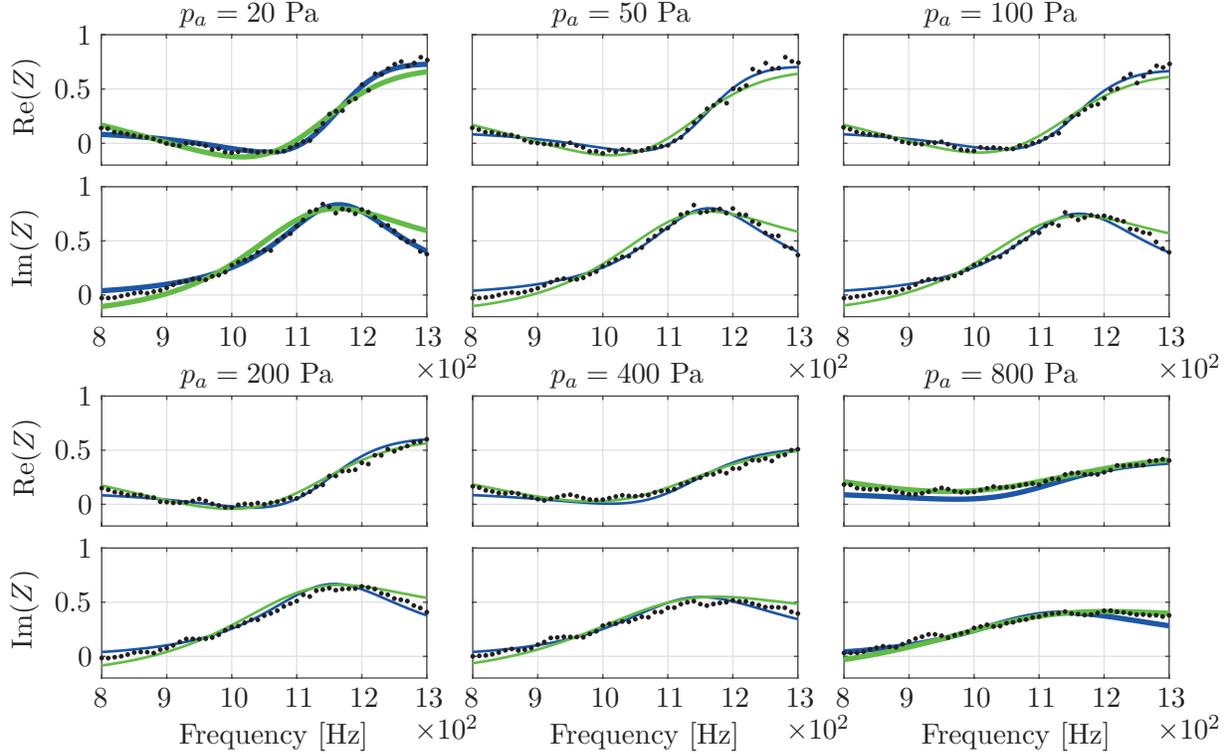}}
\end{psfrags}
  \caption{Comparison of the specific acoustic impedance $Z$ of the side branch opening, computed from model 1 (solid green line) and model 2 (solid blue line) with the experimentally measured values of the same from \cite{bourquard_faure-beaulieu_noiray_2021} (black dots). Shown are the real (first and third row) and imaginary parts (second and fourth row) of $Z$ for different acoustic pressure forcing amplitudes $p_a$ at the bulk flow velocity $U=74.1$ m/s. The model parameters were calibrated to the data for the cases $p_a=20$ Pa and $p_a=800$ Pa, as detailed in section $\S$\ref{sec:: Section 2: Modelling}. The curves for which \tr{the calibration} of the empirical parameters in the models was performed are indicated by an increased line thickness compared to the other cases.}
\label{fig:Figure 6}
\end{figure*}
\tr{\subsection{Representation of the hydrodynamic response and the saturation effect \label{Hydrodynamic response}}}
In this section, we compare the representation of the nonlinear hydrodynamic response in the side branch opening \tr{and of the saturation effect} by model 1 and 2. This response is described, for model 1, in terms of the vertical velocity induced by the vortex sheet displacement, \tr{$\mathrm{Re}(-i\omega \zeta(x)e^{-i\omega t})$}, and for model 2 by $v_{y,c}(x,t)$. In Fig. \ref{Figure 7}, we compare these two velocity fields over the aperture at a forcing frequency $f=1050$ Hz and grazing flow speed $U=74.1$ m/s for increasing forcing pressure amplitudes $p_a=20$ Pa (blue curves), $p_a=200$ Pa (red curves) and $p_a=400$ Pa (green curves) to investigate \tr{the prediction of the shear layer response} by the models from the hydrodynamic perspective. \fc{We have selected this forcing frequency because it is typical of \rv{self-sustained} aeroacoustic oscillations reported in \cite{bourquard_faure-beaulieu_noiray_2021} for a closed side branch of length $L$ ranging from $200$ to $250$ mm.} As we can see in Fig. \ref{fig:Figure 5}, for these values of $f$ and $U$, the necessary condition for instability $\mathrm{Re}(Z)<0$ is satisfied for small enough $p_a$, until $p_a$ exceeds a value of about $100$ Pa. The dashed and continuous curves in Fig. \ref{Figure 7} correspond to model 1 and 2, respectively. The phase $\omega t$ of the acoustic forcing is displayed above each frame. In the figure, we see that at $p_a=20$ Pa, there is good agreement between models 1 and 2, except for the large singular peak of the velocity field predicted by model 1 at the downstream corner, which appears at all $p_a$ and is characteristic of Howe's vortex sheet formulation \cite{howe_1998}. At the two larger forcing pressure amplitudes, there is only qualitative agreement between \tr{model 1 and 2: they roughly} agree on the position of the peak\tr{, }the propagation speed of the disturbance in the velocity field, and on the order of magnitude of this disturbance. \tr{We see from the figure that the \fc{amplitude of the coherent vertical velocity increases with the acoustic amplitude. However, as we see in eq. (\ref{Modification 1: U_-}) for model 1 and in eq. (\ref{g function 1}) for model 2, this increase is not proportional to the increase of the forcing amplitude, which, although not self-evident from this plot, leads to a saturation of the specific acoustic impedance.}}
\begin{figure*}
\begin{psfrags}
\psfrag{a}{$v_{y,c}(x,t)$ [m/s]}
\psfrag{b}{$x$ [mm]}
\psfrag{c}{$2\pi$}
\psfrag{m}{15}
\psfrag{n}{10}
\psfrag{0}{0}
\psfrag{d}{$\pi/5$}
\psfrag{e}{$2\pi/5$}
\psfrag{f}{$3\pi/5$}
\psfrag{g}{$4\pi/5$}
\psfrag{h}{$\pi$}
\psfrag{i}{$6\pi/5$}
\psfrag{j}{$7\pi/5$}
\psfrag{k}{$8\pi/5$}
\psfrag{l}{$9\pi/5$}
\psfrag{Q}{100}
\psfrag{R}{0}
\psfrag{S}{-100}
\psfrag{D}{}
\psfrag{Z}{}
\psfrag{T}{-15}
\psfrag{P}{-15}
\psfrag{L}{-15}
\psfrag{N}{15}
\psfrag{W}{0}
\psfrag{O}{15}
\psfrag{M}{$\hspace{1.24cm} 0$    }
    \centerline{\includegraphics{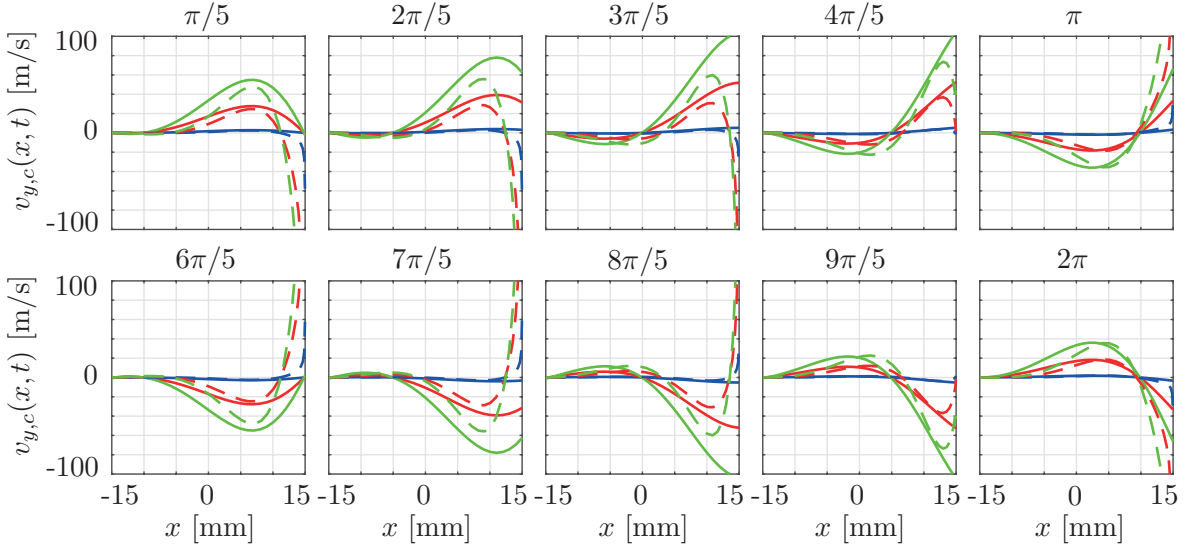}}
    \end{psfrags}
    \caption{Hydrodynamic response in the side branch opening: Shown are the vertical velocity induced by the vortex sheet displacement, $\mathrm{Re}(-i\omega \zeta(x,t))$ (dashed curves), predicted by model 1, and the velocity field $v_{y,c}(x,t)$ (continuous curves), given by model 2, in the aperture at the forcing frequency $f=1050$ Hz and bulk flow velocity $U=74.1$ m/s for increasing forcing pressure amplitudes $p_a=50$ Pa (blue curves), $p_a=200$ Pa (red curves) and $p_a=400$ Pa (green curves). The phase $\omega t$ of the acoustic forcing is displayed above each frame. }
    \label{Figure 7}
\end{figure*}

\begin{figure*}
\begin{psfrags}
\psfrag{a}{$|R|$ [-]}
\psfrag{b}{$\angle R$ [rad]}
\psfrag{c}{$1.2$}
\psfrag{d}{$1.1$}
\psfrag{e}{$1$}
\psfrag{f}{$0.9$}
\psfrag{g}{$0.8$}
\psfrag{h}{$0.7$}
\psfrag{i}{$2.6$}
\psfrag{j}{$2.55$}
\psfrag{k}{$2.5$}
\psfrag{l}{$2.45$}
\psfrag{m}{$2.4$}
\psfrag{n}{$2.35$}
\psfrag{o}{$0$}
\psfrag{p}{$200$}
\psfrag{q}{$400$}
\psfrag{s}{$600$}
\psfrag{t}{$p_a$ [Pa]}
\psfrag{u}{a}
\psfrag{v}{b}
\psfrag{A}{\tr{(a)}}
\psfrag{B}{\tr{(b)}}

    \centerline{\includegraphics{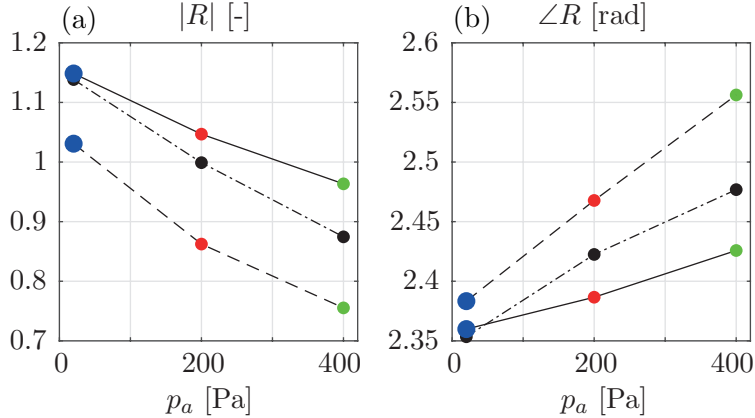}}
    \end{psfrags}
    \caption{\tr{(a)} Magnitude and \tr{(b)} phase of the reflection coefficient $R=(Z-1)/(Z+1)$ at the forcing frequency $f=1050$ Hz and grazing flow speed $U=74.1$ m/s for increasing forcing pressure amplitudes $p_a=50$ Pa (blue dots), $p_a=200$ Pa (red dots) and $p_a=400$ Pa (green dots). The dashed and continuous lines correspond to model 1 and 2, respectively.  Also shown are the experimentally measured values of $|R|$ and $\angle(R)$ at the same conditions, indicated by the black dots and the dash-dotted curves. The conditions for which \tr{the calibration} of the empirical parameters in the models was performed are indicated by an increased marker size compared to the other cases.}
    \label{Figure 8}
\end{figure*}

\tr{To compare the models' representation of the saturation effect, we show, in Fig.} \ref{Figure 8}\tr{(a),} the \tr{magnitude} and in Fig. \tr{\ref{Figure 8}\tr{(b)}} the phase of the acoustic reflection coefficient $R=(Z-1)/(Z+1)$ \tr{of the side branch aperture} at the forcing frequency $f=1050$ Hz and grazing flow speed $U=74.1$ m/s for increasing acoustic pressure forcing amplitudes $p_a=20$ Pa (blue dots), $p_a=200$ Pa (red dots) and $p_a=400$ Pa (green dots). \fc{This reflection coefficient relates the complex amplitude of the reflected wave for an incident wave originating from the side branch.} The dashed and continuous lines correspond to model 1 and 2, respectively. Also shown are the experimentally measured values of $Z$ at the same conditions, indicated by the black dots and the dash-dotted curves. \tr{The condition for which the calibration of the} parameters in the models was performed is indicated by an increased marker size compared to the other cases. In the figure, for both models \rv{as well as} for the values from the experiments, we see a decline in the acoustic gain $|R|$ from values above $1$ at low forcing amplitudes $p_a$ to values below $1$ at high $p_a$ and a small increase in the phase $\angle(R)$ as $p_a$ is increased. \tr{These analytical predictions of $R$ are in agreement with the results obtained in \cite{boujo_bauerheim_noiray_2018} using compressible LES and incompressible LNSE analysis and with the experiments of \cite{bourquard_faure-beaulieu_noiray_2021}}. We note that model 2 compares better than model 1 to the experimentally measured values of the reflection coefficient $R$. The discrepancies between models 1 and 2 seen in Fig. \ref{Figure 8}, which may seem surprising in light of the good overall agreement between the two models shown in Fig. \ref{fig:Figure 6}, can be explained by the fact that the empirical parameters were calibrated to data acquired over a large frequency \tr{range}, and a good global fit of $Z$ in this domain does not necessarily imply similarly good agreement \tr{of $Z$ or $R$} between the models and the experiments \tr{at each frequency}.

\begin{figure*}
\begin{psfrags}
\psfrag{a}{$200$}
\psfrag{b}{$400$}
\psfrag{c}{$600$}
\psfrag{d}{$900$}
\psfrag{e}{$1050$}
\psfrag{f}{$1200$}
\psfrag{g}{$1200$}
\psfrag{h}{$0$}
\psfrag{i}{$0.5$}
\psfrag{j}{$1$}
\psfrag{k}{$20$}
\psfrag{l}{$800$}
\psfrag{m}{$|R|$}
\psfrag{n}{Model 1}
\psfrag{o}{Model 2}
\psfrag{p}{$f$ [Hz]}
\psfrag{s}{$p_a$ [Pa]}
\psfrag{t}{$p_a$ [Pa]}
\psfrag{A}{$63.5$ m/s}
\psfrag{B}{$67.1$ m/s}
\psfrag{C}{$70.6$ m/s}
\psfrag{D}{$74.1$ m/s}
\psfrag{E}{}
\psfrag{F}{}
\psfrag{G}{$U$}
\centerline{\includegraphics{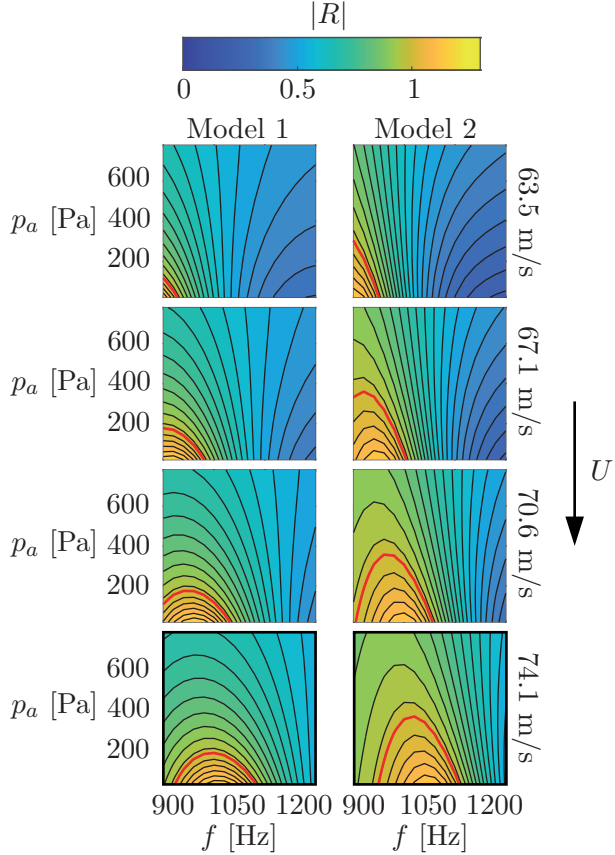}}
    \end{psfrags}
    \caption{\tr{Visualization of the saturation effect over a range of frequencies and acoustic pressure forcing amplitudes $p_a$ for increasing bulk flow velocities $U$. Shown are the contour plots of $|R|$ for different values of $U$. The black arrow indicates the direction of increasing $U$. The condition for which the calibration of the parameters in the models was performed is indicated by a black frame around the respective insets.}}
    \label{Figure 9}
\end{figure*}

\tr{In Fig. \ref{Figure 9}, we visualize the saturation effect predicted by the models with contour plots of the magnitude of the reflection coefficient of the side branch aperture $|R|$ over a range of frequencies and acoustic pressure forcing amplitudes $p_a$ for increasing bulk flow velocities $U$. The black arrow indicates the direction of increasing $U$. The contour for which $|R|=1$ is indicated by a red curve with increased thickness. The condition for which the calibration of the parameters in the models was performed is indicated by a black frame around the respective insets. The figure shows that both models agree qualitatively in their representation of the saturation effect. Model 1 predicts that the necessary condition for instability $\mathrm{Re}(Z)<0$ is satisfied in a smaller region than predicted by model 2.}
\\

\subsection{Discussion \label{sec:: Section 4: Discussion}}
The two models we have derived and analysed in $\S$\ref{sec:: Section 2: Modelling} and $\S$\ref{sec:: Section 3: Results}, respectively, include empirical parameters that require calibration to experimental data. The choice of the number of parameters we introduce is not unique and the models themselves are not the only models that can be used to achieve similar results in predicting the impedance of the side branch opening. We demonstrate, however, in appendix \ref{Appendix C}, that from the parameters in both models 1 and 2, none can be removed while still achieving a good fit to the impedance curves at grazing flow speed $U=74.1$ m/s and acoustic pressure forcing amplitude $p_a=20$ Pa. 

An alternative model that could be used instead of model 1 is the analytical model of Howe for the Rayleigh conductivity of a rectangular aperture, with streamwise and spanwise extensions $W$ and \tr{$H$}, respectively. This model is given by eq. (3.3) in \cite{Howe1996}. By adjusting the velocities above and below the shear layer, and adding two offsets as was done in model 1, similar results for the prediction of the specific acoustic impedance $Z$ can be achieved. For this model, the \fc{vortex sheet} displacement can be computed from eq. (3.2) in \cite{Howe1996}. 

Model 2 involves 8 empirical parameters, 6 of which are necessary \tr{for calibration to experimentally measured values of $Z$} at a small acoustic pressure forcing amplitude $p_a$. While it is possible, by defining the function $g$ in a different way, to derive simpler models which include less empirical parameters, we found no \tr{alternative to model 2} that could achieve \rv{similarly accurate and robust} predictions of the measured impedance curves. \rv{Note that to obtain an explicit analytical expression for $Z$ with model 2, the choice of $g$ is restricted to standard functions where the Laplace transform can be explicitly computed. Alternative $g$ functions that were tried include real lower-order polynomials (up to quartic), as well as the exponential function $\mathrm{exp}$ with argument $\varkappa x$, where $\varkappa\in\mathbb{R}$ is a real constant. With none of these alternatives, a comparably good fit to the experimental results was achieved as with the expression given in eq. \eqref{g function 1}. From a practical perspective, this led to our choice of the function $g$. The question remains open whether it is possible to make a more systematic choice of the $g$, possibly reducing the number of parameters, while maintaining the predictive quality of model 2. This question could be addressed in future research.}

\rv{The authors acknowledge the following drawback of model 2: due to the necessity of a multitude of calibrated terms, no a priori estimates of the reflection coeffient of the aperture are possible. This is to be expected, because model 2 is based on an ad hoc assumed coherent velocity profile, and is neither derived from first principles nor takes into account the problem geometry. On the other hand, model 1, which is based on the linearized unsteady Bernoulli equation and Green's functions for the side-branch geometry, does allow for such an a priori estimate, which can be obtained by setting $U_-=U/2$, $U_+=0$, $\gamma_1=0$ and $\delta=0$. With this choice, model 1 scales with the Strouhal number $\mathrm{St}=\omega W/2 \pi U$ and can be used to easily assess changes in the unstable frequency range with the mean flow velocity and the aperture width. This approximation is simply Howe's original model which is introduced in \cite{howe_1998}, pp. 446-448, example 4. From Fig. 6.1.9. therein, one sees that the maximum of the reflection coefficient will be around $f W/U_-\approx0.75-0.8$, with the exact value of the right-hand side depending on the ratio of the channel height to the aperture width. This gives, for the critical frequency of the aperture, $f\approx \{0.75-0.8\}U/2W$. Around this frequency, the amplification of the acoustic field in the cavity by the shear layer is maximal. If this critical frequency coincides with a resonance frequency of an acoustic mode of the cavity and if the cavity damping is small enough, an aeroacoustic instability can occur. The approximation $U_-=U/2$ is further discussed in the last paragraph of appendix \ref{Appendix C}. Note that for $\delta\neq0$, model 1 does not scale with the Strouhal number. Replacing the correction term $i \omega \delta $ with, e.g., $i \omega \delta /U$ or a constant imaginary term $i\delta$ yielded overall worse results over the relevant frequency range than the present imaginary correction term, which does not have Strouhal scaling. Note that in contrast, without any further modification, model 2 does scale with the Strouhal number.} 

\rv{Furthermore, we note that in Fig. \ref{fig:Figure 5}, the reactance maximum is overpredicted by both models, with starker contrast at lower grazing flow speeds. For model 1, this could be rectified by replacing the correction term $i \omega \delta $ with $i \omega W\delta /U$. However, although the peak of the reactance is better predicted with this term, as stated above, the overall match between model 1 and experiments becomes worse. For model 2, it is not clear how this shortcoming could be improved.} 

By prescribing a different \tr{coherent displacement field} $\tilde{y}(x,t)$ altogether \rv{(as opposed to simply replacing the function $g$)}, alternative models can be derived along the same lines as model 2. This includes models that involve the interaction of two separate hydrodynamic modes of the shear layer. \tr{The interested reader can refer to the stability analysis presented in \cite{boujo_bauerheim_noiray_2018} using incompressible LNSE, which shows that there are, in the side branch opening geometry, two shear layer modes around the Strouhal numbers $\mathrm{St}=\omega W/2\pi U=0.4$ and $\mathrm{St}=0.53$.} Such an interaction could not be captured with model 2, which is acceptable for the purpose of this work, where we are concerned with describing the acoustic response of the side branch opening in the vicinity of the eigenfrequency of $\it{one}$ of the hydrodynamic modes in the \tr{side branch} aperture.

\tr{Due the lack of PIV data in the relevant frequency range, comparison between such data and the predictions of the hydrodynamic response in the side branch aperture by model 1 and 2 was not possible. However, we note that the limiting assumptions of idealized models such as 3D side wall effects on the eigenmodes will affect the convective speed of perturbations in the aperture and therefore the Strouhal number of the maximum of $|R|$, \rv{as well as} the decay rate of these hydrodynamic modes and therefore the value of the maximal $|R|$. This is in particular why in model 2, $u_c$ substantially differs from an educated guess of $U/2$ for an ideal 2D side branch. For these reasons, even if PIV data had been available in the frequency range of interest, i.e., in the vicinity of the eigenfrequency of the first hydrodynamic eigenmode of the aperture, a comparison between this data and the hydrodynamic response predicted by model 1 and 2 still would not be straightforward and disagreements would be expected due to these 3D effects.} 

\section{Conclusions \label{sec:: Section 4: Conclusions}}
We have derived two models for the acoustic impedance, which characterizes the aeroacoustic response of the opening of a side-branch cavity subject to harmonic acoustic pressure forcing under turbulent grazing flow. We showed that, after calibration to \tr{experimental} impedance data, these models \tr{robustly predict} the measured impedance curves for a broad range of the frequency, bulk flow velocity and acoustic pressure \rv{forcing amplitude}. The aeroacoustic response of the side branch opening is one element of the classic transfer function formalism that is widely used to describe and predict self-sustained cavity oscillations, a phenomenon that is relevant in many industrial applications. Hence the models developed in this work can serve as parts of predictive network models that aim to quantitatively describe self-sustained \tr{aeroacoustic} oscillations \tr{in cavities} for different \rv{bulk} flow \tr{speeds}. We compared the models' representations of the hydrodynamic response in the side branch aperture \tr{and of the saturation effect} for increasing acoustic pressure forcing amplitudes $p_a$, showing qualitative agreement between the two models. A comparison of this hydrodynamic response to PIV data was not possible due to a lack of data in the relevant frequency range, but is a topic for future research. \tr{Alternative models to the ones presented in this work were also discussed}.

\section*{Acknowledgements}
The authors are grateful to Michael Howe for helpful \tr{discussion} about his previous work. This project is funded by the Swiss National Science Foundation under Grant agreement 184617, and by the European Union’s Horizon 2020 Research and Innovation Programme under Grant Agreement No. 765998.

\appendix
\section{}\label{Appendix A}
\tr{In this section, we provide elements of the derivation from \citet{howe_1998} for eq. (\ref{Master equation for zeta (Section 2)}) which constitutes the basis of model 1. For this derivation, we write the acoustic pressure load as $p_a=p_--p_+$, where $p_+$ and $p_-$ are the uniform components of the pressure on both sides of the vortex sheet in the side branch aperture. This derivation is based on a linear approximation of the vortex sheet, the \rv{linearized unsteady Bernoulli equation and pressure continuity} at the side branch opening. The latter two ingredients lead to 
\begin{equation}
    p_++i\rho_0\left(\omega+iU_+\frac{\partial}{\partial x}\right)\phi_+=p_-+i\rho_0\left(\omega+iU_-\frac{\partial}{\partial x}\right)\phi_-, \label{A1}
\end{equation}
where $U_+$ and $U_-$ are the streamwise mean velocities just above and below the vortex sheet, and $\phi_\pm$ is the velocity potential that is \fc{associated with} the vertical velocity component which satisfies the boundary conditions of the rectangular duct and \rv{the} side branch. Considering that this vertical velocity at the side branch opening is linked to the vertical displacement of the vortex sheet $\zeta$ via $v_{y,\pm}=\left(\frac{\partial}{\partial t} +U_\pm \frac{\partial}{\partial x}\right) \zeta=-i\left(\omega+i U_\pm \frac{\partial}{\partial x}\right)\zeta$, and nondimensionalizing time and spatial coordinates, Howe obtains the following expression for the left and right hand sides \fc{(LHS and RHS, respectively)} of eq. (\ref{A1}). First, the \fc{RHS} of this equation becomes
\begin{equation}
    p_--\frac{2\rho_0 U_-^2 }{\pi W }\left(\sigma+i\frac{\partial}{\partial\xi}\right)^{2} \int_{-1}^1\zeta(\mu)\left(\mathrm{ln}|\xi-\mu|+L_-(\xi,\mu)\right)d\mu,
\end{equation}
where $\mu$ is an integration variable corresponding to $\xi=2x/W$, $\sigma=\omega W/2U_-$ is the Kelvin-Helmholtz wavenumber and
\begin{equation}
    \rv{L_-(\xi,\mu)=\mathrm{ln}\left(\frac{2 \hspace{0.1mm}\mathrm{sinh} \{ \pi W(\xi-\mu)/4H \} }{\xi-\mu}\right).}
\end{equation} 
This expression can be found in \cite{howe_1998}, p. 447. Second, the \fc{LHS} of eq. (\ref{A1}) is \begin{equation}
    p_++\frac{2\rho_0 U_-^2 }{\pi W}\left( \sigma+i \alpha \frac{\partial}{\partial\xi}\right)^2 \int_{-1}^1\zeta(\mu)\left(\mathrm{ln}|\xi-\mu|+L_+(\xi,\mu)\right)d\mu,
\end{equation}
where 
\begin{equation}
    L_+(\xi,\mu)=\mathrm{ln}\left(\frac{4\sin{ \{ \pi(\xi-\mu)/4 \}}\cos{ \{ \pi(\xi+\mu)/4 \}} }{\xi-\mu}\right)
\end{equation}
\fc{and $\alpha=U_+/U_-$.} This expression differs from that given in \cite{howe_1998}, pp. 445-446, because we consider a nonzero streamwise velocity above the vortex sheet. From the above equations, one obtains
\begin{eqnarray}
&&\left[\left( \sigma+i \alpha \frac{\partial}{\partial\xi}\right)^2+\left(\sigma+i\frac{\partial}{\partial\xi}\right)^{2}\right]\int_{-1}^{1}\zeta(\mu)\{\mathrm{ln}|\xi-\mu|+L_{-}(\xi,\mu)\}d\mu\nonumber\\
&&+\left( \sigma+i \alpha \frac{\partial}{\partial\xi}\right)^2\int_{-1}^{1}\zeta(\mu)\{L_{+}(\xi,\mu)-L_{-}(\xi,\mu)\}d\mu= \frac{\pi W p_a }{2 \rho_0 U_-^2}\nonumber\\
&&\approx\left[\left( \sigma+i \alpha \frac{\partial}{\partial\xi}\right)^2+\left(\sigma+i\frac{\partial}{\partial\xi}\right)^{2}\right]\int_{-1}^{1}\zeta(\mu)\{\mathrm{ln}|\xi-\mu|+L_{-}(\xi,\mu)\}d\mu \nonumber\\
&&+\sigma^{2}\int_{-1}^{1}\zeta(\mu)\{L_{+}(\xi,\mu)-L_{-}(\xi,\mu)\}d\mu= \frac{\pi W p_a }{2 \rho_0 U_-^2},
\label{Diff. eq. for zeta}
\end{eqnarray}
where the term $i\alpha\frac{\partial}{\partial \xi}$ was neglected in the bracket before the second integral. This is justified by the following considerations. First, we can write the respective term as
\begin{equation}
    \left( \sigma+i \alpha \frac{\partial}{\partial\xi}\right)^2\int_{-1}^{1}\zeta(\mu)\{L_{+}(\xi,\mu)-L_{-}(\xi,\mu)\}d\mu. \label{Howe term}
\end{equation}
In Howe's theory, $\zeta\equiv 0$ \fc{outside} the aperture. By the symmetry of $L_+$ and $L_-$ in their arguments and partial integration, we can rewrite eq. (\ref{Howe term}) as
\begin{equation}
    \int_{-1}^{1}\left[\left( \sigma+i \alpha \frac{\partial}{\partial\mu}\right)^2\zeta(\mu)\right]\{L_{+}(\xi,\mu)-L_{-}(\xi,\mu)\}d\mu. 
\end{equation}
The factor $\alpha$ is assumed to be small and since it multiplies only bounded terms, the error in the solution $\zeta$ we incur from dropping these terms will be \fc{of order $O(\alpha)$, i.e., small}. This simplification enables the following analytical manipulations. To integrate eq. (\ref{Diff. eq. for zeta}), we note that the Green's function for the operator
\begin{equation}
\left( \sigma+i \alpha \frac{\partial}{\partial\xi}\right)^2+\left(\sigma+i\frac{\partial}{\partial\xi}\right)^{2} \label{2nd order diff. operator}
\end{equation}
is given by eq. (6.1.7) in \cite{howe_1998}:
\begin{equation}
 G(\xi,\mu)=\frac{1}{2\sigma(1-\alpha)}\left( H(\xi-\mu)e^{i\sigma_1(\xi-\mu)}+H(\mu-\xi)e^{i\sigma_2(\xi-\mu)}\right),
\end{equation}
where $H(x)$ is the Heaviside function and
\begin{equation}
    \sigma_{1,2}=\sigma \frac{1\pm i}{1 \pm i \alpha}.
\end{equation}
We further note that the kernel of the operator (\ref{2nd order diff. operator}) is given by $\lambda_1 e^{i \sigma_1 \xi}+\lambda_2 e^{i \sigma_2 \xi}$. Hence we can rewrite eq. (\ref{Diff. eq. for zeta}) as follows:
\begin{equation}
    \int_{-1}^1 \zeta'(\mu) \left[\ln{|\xi-\mu|}+L_-(\xi,\mu)+K(\xi,\mu)\right] d\mu+\lambda_1 e^{i \sigma_1 \xi}+\lambda_2 e^{i \sigma_2 \xi}=1, \label{Master equation for zeta}
\end{equation}
where we have defined $\zeta'=\zeta\rho_0\omega^2 W/\pi p_a$ and 
\begin{eqnarray}
K(\xi,\mu)=&&\frac{\sigma}{2(1-\alpha)}\int_{-1}^1 \{L_+(\lambda,\mu)-L_-(\lambda,\mu)\} \nonumber\\
&&\times\left( H(\xi-\lambda)e^{i\sigma_1(\xi-\lambda)}+H(\lambda-\xi)e^{i\sigma_2(\xi-\lambda)}\right) d\lambda. \label{Modified K}
\end{eqnarray}
The solution $\zeta(\xi)$ of (\ref{Master equation for zeta}) which satisfies the Kutta condition \begin{equation}
    \zeta'(-1)=\frac{\partial \zeta'}{\partial \xi}(-1)=0, \label{Kutta condition, Appendix}
\end{equation} 
 constitutes the basis of model 1. The method that was used to solve eq. (\ref{Master equation for zeta}) is detailed in appendix \ref{Appendix B}.}

\section{}\label{Appendix B}
In this section, we describe a solution method for linear integral equations of the form 
\begin{equation}
\int_{a}^{b}F(x,y)g(y)dy+h(x)=0,\label{eq:Linear integral equation}
\end{equation}
with given complex-valued functions $F(x,y)$ and $h(x)$ for an unknown complex-valued function $g(x)$ on the domain $\left[a,b\right]$. We achieve this using Gauss-Legendre quadrature of order $N$ \tr{\cite{abramowitz1948handbook}}. This rule transforms an integral 
\begin{equation}
\tr{\int_{a}^{b}g(y)dy}
\end{equation}
into a sum of weights and function values of $g$, evaluated at the
points $y_{i}$ for \tr{$i=1,...,N$}:
\begin{equation}
\tr{\int_{a}^{b}g(y)dy}=\sum_{i=1}^{N}g(y_{i})w(y_{i}).\label{eq: Gauss-Legendre rule}
\end{equation}
\tr{The points $y_i$ are defined, see eq. (25.4.30) in \cite{abramowitz1948handbook}, as $y_i=(b-a)x_i/2+(b+a)/2$ and the weights $w(y_i)$ as $w(y_i)=(b-a)(P_N'(x_i))^2/(1-x_i^2)$, where $P_N$ is the Legendre polynomial of order $N$ \cite{abramowitz1948handbook} and $x_i$ is the $i^{\mathrm{th}}$ zero of $P_N$.} Note that eq. (\ref{eq:Linear integral equation}) needs to be
satisfied at all points $x\in[a,b].$ Hence \tr{for any point $x_j$ in this interval, we can} rewrite \tr{eq.} (\ref{eq:Linear integral equation})
using (\ref{eq: Gauss-Legendre rule}) \tr{as}
\tr{\begin{equation}
\int_{a}^{b}F(x_{j},y)g(y)dy+\tr{h}(x_{j})=\sum_{i=1}^{N}F(x_{j},y_{i})g(y_{i})w(y_{i})+h(x_{j})=0.\label{eq:Linear integral equation with Gauss rule at x1}
\end{equation}}
Doing so for $M$ points $x_{j}=a,...,b$ gives a linear system of $M$ equations
\begin{equation}
\sum_{i=1}^{N}F(x_{j},y_{i})g(y_{i})w(y_{i})+h(x_{j})=0,
\end{equation}
which can be written as 
\begin{equation}
Az=c,\label{eq: Linear system of equations}
\end{equation}
where $A\in\mathbb{R}^{^{M\times N}}$, $A_{ij}=F(x_{j},y_{i})w(y_{i})$, $z\in\mathbb{R}^M$, $z_{i}=g(y_{i})$ for $i=1,...,N$
and $c\in\mathbb{R}^M$, $c_{j}=-h(x_{j})$ for $j=1,...,M.$ The linear system of equations
(\ref{eq: Linear system of equations}) is then solved for $z$ by a covariance matrix adaptation evolution strategy (CMAES) algorithm (\citet{CMAES}) which minimizes $||Az-c||$. As an initial guess, the
least squares solution of (\ref{eq: Linear system of equations})
was used. Gauss-Legendre quadrature of order $10$ was used in this paper to solve the linear integral equation (\ref{Master equation for zeta}) to obtain the acoustic impedance, and the same method of order $40$ was used to obtain the velocity field induced vortex sheet displacement \tr{shown in Fig. \ref{Figure 7} in} $\S$\ref{sec:: Section 3: Results}. \tr{This high order was chosen because a sharp resolution of $\zeta$ was required for visualization and the CMAES algorithm is more efficient if the matrix $A$ in eq. (\ref{eq: Linear system of equations}) is nearly quadratic.} \tr{The Kutta condition (\ref{Kutta condition, Appendix}) can be implemented by setting $\zeta(\xi_1)=\zeta(\xi_2)=0$, following the remarks in \cite{howe_1998}, p. 436.} We note that care must be taken when numerically evaluating integrals of functions which have a singularity on the interior of the integration domain as it is the case, e.g., with \tr{the function $K$ given by eq.} (\ref{Modified K}). In this case, the integral must be decomposed into multiple integrals so that all singularities lie on a boundary. The computational method used to solve eq. (\ref{Master equation for zeta}) was benchmarked using the results shown in Fig. 6.1.9. in \cite{howe_1998}. 

\section{}\label{Appendix C}
\begin{figure*}
\begin{psfrags}
\psfrag{a}{$1$}
\psfrag{b}{$0.5$}
\psfrag{c}{$0$}
\psfrag{h}{$2$}
\psfrag{f}{$4$}
\psfrag{p}{$4$}
\psfrag{n}{$8$}
\psfrag{l}{$12$}
\psfrag{A}{\tr{(a)}}
\psfrag{B}{\tr{(b)}}
\psfrag{C}{\tr{(d)}}
\psfrag{F}{\tr{(c)}}
\psfrag{D}{\tr{(e)}}
\psfrag{L}{\tr{(f)}}
\psfrag{K}{$800$}
\psfrag{E}{$1000$}
\psfrag{G}{$1200$}
\psfrag{Z}{Frequency [Hz]}
\psfrag{k}{$S$}
\psfrag{Y}{$\mathrm{Re}(Z)$}
\psfrag{W}{$\mathrm{Im}(Z)$}
\psfrag{U}{Model 1}
\psfrag{X}{Model 2}
\psfrag{d}{$\alpha_0$\hspace{.15cm}$\gamma$\hspace{.27cm}$\beta$\hspace{.245cm}$\delta$}
\psfrag{I}{\hspace{0.07cm}$\gamma$\hspace{.12cm}$a_3$\hspace{.07cm}$a_4$\hspace{.06cm}$a_5$\hspace{.065cm}$a_2$}

    \centerline{\includegraphics{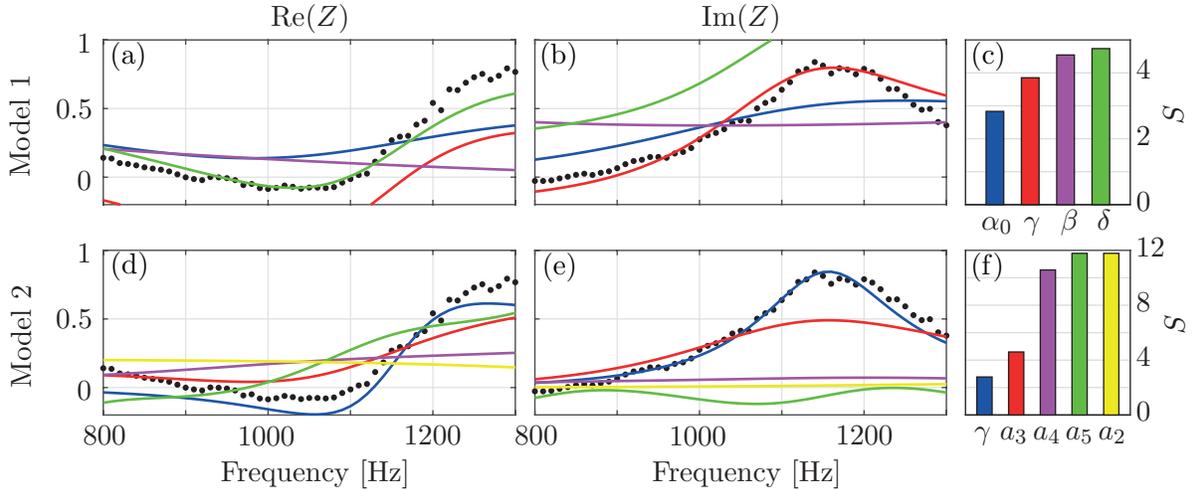}}
    \end{psfrags}
    \caption{Sensitivity $S$, defined in \rv{eq.} \eqref{Relative RMSD}, for models 1 and 2 at $U=74.1$ m/s and $p_a=20$ Pa with respect to the empirical parameters shown in the $x$-axis. The parameters are ordered by increasing $S$. The bar plot shows the relative increase of RMSD between the model and the experimentally acquired values of $Z$ after calibration when a given empirical parameter is excluded from the respective model.}
    \label{Figure 10}
\end{figure*}
In this section, we investigate the sensitivity of models 1 and 2 with respect to the empirical parameters that were included in these models to obtain predictions for different grazing flow speeds $U$. To determine the sensitivity of models 1 and 2 with respect to a parameter for given $U$ and $p_a$, we compute the root-mean-square deviation ($\mathrm{RMSD}$) between the model and the experimental data, defined as
\begin{equation}
    \mathrm{RMSD}=\sqrt{\frac{1}{N_f}\sum_{i=1}^{N_f} {|Z^\mathrm{Model}_i-Z^\mathrm{Experiment}_i|}^2},
\end{equation}
where $N_f$ is the number of frequency points at which $Z$ was acquired. In the present case, $N_f=51$. We denote the value of $\mathrm{RMSD}$ of the full models including all parameters by $\mathrm{RMSD}_0$. To measure a model's sensitivity \rv{to its parameters}, we exclude each parameter, creating a reduced model with one less parameter and perform a fit of this reduced model to the impedance curves. As starting values for the fit of the remaining parameters, we used the respective values obtained from calibrating the full model. We then compute $\mathrm{RMSD}$ for each reduced model and compute the relative value compared to the full model, which we define as the sensitivity $S$ of this parameter:
\begin{equation}
    S=\mathrm{RMSD}/\mathrm{RMSD}_0. \label{Relative RMSD}
\end{equation}
The values of $S$ are shown in Fig. \ref{Figure 10}\tr{(c)} and \tr{\ref{Figure 10}(f)} for model 1 and 2 at $U=74.1$ m/s and $p_a=20$ Pa with respect to the empirical parameters shown on the $x$-axis. As shown by the thick curves in Fig. \ref{fig:Figure 3}, for these values of $U$ and $p_a$, a good fit was \tr{achieved} with both models 1 and 2 over the considered frequency range. In Fig. \ref{Figure 10}, the parameters are ordered by increasing $S$. The bar plots in Fig. \ref{Figure 10}(c) and \ref{Figure 10}(f) show the relative increase of RMSD between the models and the experimentally acquired values of $Z$ after calibration when a given empirical parameter is excluded, i.e., set equal to zero in the respective model. An exception is the parameter $\beta$, which we removed from model 1 by setting it equal to $1$, so that \rv{$U_-=U$}. We see that for both models, removing any one of the parameters leads to an increase in $\mathrm{RMSD}$ of at least a factor 2. In Fig. \ref{Figure 10}(a) and \ref{Figure 10}(b) and \ref{Figure 10}(d) and \ref{Figure 10}(e), we show the real and imaginary parts \rv{of $Z$}, respectively, over the considered frequency \tr{range} after the fit of the reduced models, showing that the achieved fit is visibly bad with after removing any of the parameters in the models. The effect of removing the parameter $\kappa$ from model 2, i.e., setting it equal to $1$, is similar to removing $\beta$ from model 1: It does not allow the impedance curves to be squeezed in the frequency domain, so that the undulated portion lies in the range of the experimental results. The corresponding bar plot and impedance curves are not shown in Fig. \ref{Figure 10} for the sake of a compact presentation. 

\rv{We want to highlight that the above results show that, in a first approximation, the mean flow speed just below the vortex sheet $U_-$ should not be set equal to the total mean flow speed $U$, but to $U/2$. This approximation accounts for the sharp drop-off of the mean flow speed in the presence of the turbulent boundary layer at the wall. Due to this drop-off, vorticity fluctuations in the aperture are advected at a significantly lower speed than the mean flow speed $U$ away from the wall. Using this approximation, without any correction terms and $U_+=0$, model 1, in its original form given by \citet{howe_1998}, can be used as an a priori tool to estimate the reflection coefficient $|R|$ of the aperture at a given condition. Such an a priori estimate is not possible with model 2, which is based on an ad hoc assumed coherent velocity field profile in the aperture that is calibrated to experiments.}

\bibliography{apssamp}

\providecommand{\noopsort}[1]{}\providecommand{\singleletter}[1]{#1}%
\begin{thebibliography}{31}%
\makeatletter
\providecommand \@ifxundefined [1]{%
 \@ifx{#1\undefined}
}%
\providecommand \@ifnum [1]{%
 \ifnum #1\expandafter \@firstoftwo
 \else \expandafter \@secondoftwo
 \fi
}%
\providecommand \@ifx [1]{%
 \ifx #1\expandafter \@firstoftwo
 \else \expandafter \@secondoftwo
 \fi
}%
\providecommand \natexlab [1]{#1}%
\providecommand \enquote  [1]{``#1''}%
\providecommand \bibnamefont  [1]{#1}%
\providecommand \bibfnamefont [1]{#1}%
\providecommand \citenamefont [1]{#1}%
\providecommand \href@noop [0]{\@secondoftwo}%
\providecommand \href [0]{\begingroup \@sanitize@url \@href}%
\providecommand \@href[1]{\@@startlink{#1}\@@href}%
\providecommand \@@href[1]{\endgroup#1\@@endlink}%
\providecommand \@sanitize@url [0]{\catcode `\\12\catcode `\$12\catcode
  `\&12\catcode `\#12\catcode `\^12\catcode `\_12\catcode `\%12\relax}%
\providecommand \@@startlink[1]{}%
\providecommand \@@endlink[0]{}%
\providecommand \url  [0]{\begingroup\@sanitize@url \@url }%
\providecommand \@url [1]{\endgroup\@href {#1}{\urlprefix }}%
\providecommand \urlprefix  [0]{URL }%
\providecommand \Eprint [0]{\href }%
\providecommand \doibase [0]{https://doi.org/}%
\providecommand \selectlanguage [0]{\@gobble}%
\providecommand \bibinfo  [0]{\@secondoftwo}%
\providecommand \bibfield  [0]{\@secondoftwo}%
\providecommand \translation [1]{[#1]}%
\providecommand \BibitemOpen [0]{}%
\providecommand \bibitemStop [0]{}%
\providecommand \bibitemNoStop [0]{.\EOS\space}%
\providecommand \EOS [0]{\spacefactor3000\relax}%
\providecommand \BibitemShut  [1]{\csname bibitem#1\endcsname}%
\let\auto@bib@innerbib\@empty
\bibitem [{\citenamefont {Sondhauss}(1854)}]{sondhauss1854ueber}%
  \BibitemOpen
  \bibfield  {author} {\bibinfo {author} {\bibfnamefont {C.}~\bibnamefont
  {Sondhauss}},\ }\bibfield  {title} {\bibinfo {title} {{\"U}ber die beim
  {A}usstr{\"o}men der {L}uft entstehenden {T}{\"o}ne},\ }\href@noop {}
  {\bibfield  {journal} {\bibinfo  {journal} {Ann. Phys. (Leipzig)}\ }\textbf
  {\bibinfo {volume} {167}},\ \bibinfo {pages} {214} (\bibinfo {year}
  {1854})}\BibitemShut {NoStop}%
\bibitem [{\citenamefont {Wilson}\ \emph {et~al.}(1971)\citenamefont {Wilson},
  \citenamefont {Beavers}, \citenamefont {DeCoster}, \citenamefont {Holger},\
  and\ \citenamefont {Regenfuss}}]{wilson1971experiments}%
  \BibitemOpen
  \bibfield  {author} {\bibinfo {author} {\bibfnamefont {T.}~\bibnamefont
  {Wilson}}, \bibinfo {author} {\bibfnamefont {G.}~\bibnamefont {Beavers}},
  \bibinfo {author} {\bibfnamefont {M.}~\bibnamefont {DeCoster}}, \bibinfo
  {author} {\bibfnamefont {D.}~\bibnamefont {Holger}},\ and\ \bibinfo {author}
  {\bibfnamefont {M.}~\bibnamefont {Regenfuss}},\ }\bibfield  {title} {\bibinfo
  {title} {Experiments on the fluid mechanics of whistling},\ }\href@noop {}
  {\bibfield  {journal} {\bibinfo  {journal} {J. Acoust. Soc. Am.}\ }\textbf
  {\bibinfo {volume} {50}},\ \bibinfo {pages} {366} (\bibinfo {year}
  {1971})}\BibitemShut {NoStop}%
\bibitem [{\citenamefont {Fabre}\ \emph {et~al.}(2012)\citenamefont {Fabre},
  \citenamefont {Gilbert}, \citenamefont {Hirschberg},\ and\ \citenamefont
  {Pelorson}}]{fabre2012aeroacoustics}%
  \BibitemOpen
  \bibfield  {author} {\bibinfo {author} {\bibfnamefont {B.}~\bibnamefont
  {Fabre}}, \bibinfo {author} {\bibfnamefont {J.}~\bibnamefont {Gilbert}},
  \bibinfo {author} {\bibfnamefont {A.}~\bibnamefont {Hirschberg}},\ and\
  \bibinfo {author} {\bibfnamefont {X.}~\bibnamefont {Pelorson}},\ }\bibfield
  {title} {\bibinfo {title} {Aeroacoustics of musical instruments},\
  }\href@noop {} {\bibfield  {journal} {\bibinfo  {journal} {Annu. Rev. Fluid
  Mech.}\ }\textbf {\bibinfo {volume} {44}},\ \bibinfo {pages} {1} (\bibinfo
  {year} {2012})}\BibitemShut {NoStop}%
\bibitem [{\citenamefont {Rockwell}\ and\ \citenamefont
  {Naudascher}(1979)}]{rockwell1979self}%
  \BibitemOpen
  \bibfield  {author} {\bibinfo {author} {\bibfnamefont {D.}~\bibnamefont
  {Rockwell}}\ and\ \bibinfo {author} {\bibfnamefont {E.}~\bibnamefont
  {Naudascher}},\ }\bibfield  {title} {\bibinfo {title} {Self-sustained
  oscillations of impinging free shear layers},\ }\href@noop {} {\bibfield
  {journal} {\bibinfo  {journal} {Annu. Rev. Fluid Mech.}\ }\textbf {\bibinfo
  {volume} {11}},\ \bibinfo {pages} {67} (\bibinfo {year} {1979})}\BibitemShut
  {NoStop}%
\bibitem [{\citenamefont {Ziada}\ and\ \citenamefont
  {Lafon}(2014)}]{ziada2014flow}%
  \BibitemOpen
  \bibfield  {author} {\bibinfo {author} {\bibfnamefont {S.}~\bibnamefont
  {Ziada}}\ and\ \bibinfo {author} {\bibfnamefont {P.}~\bibnamefont {Lafon}},\
  }\bibfield  {title} {\bibinfo {title} {Flow-excited acoustic resonance
  excitation mechanism, design guidelines, and counter measures},\ }\href@noop
  {} {\bibfield  {journal} {\bibinfo  {journal} {Appl. Mech. Rev.}\ }\textbf
  {\bibinfo {volume} {66}} (\bibinfo {year} {2014})}\BibitemShut {NoStop}%
\bibitem [{\citenamefont {Rockwell}\ and\ \citenamefont
  {Naudascher}(1978)}]{rockwell1978review}%
  \BibitemOpen
  \bibfield  {author} {\bibinfo {author} {\bibfnamefont {D.}~\bibnamefont
  {Rockwell}}\ and\ \bibinfo {author} {\bibfnamefont {E.}~\bibnamefont
  {Naudascher}},\ }\bibfield  {title} {\bibinfo {title} {Review--self-sustained
  \rv{oscillations} of flow past cavities},\ }\href@noop {} {\bibfield
  {journal} {\bibinfo  {journal} {J. Fluids Eng.}\ }\textbf {\bibinfo {volume}
  {100}},\ \bibinfo {pages} {152} (\bibinfo {year} {1978})}\BibitemShut
  {NoStop}%
\bibitem [{\citenamefont {Rowley}\ and\ \citenamefont
  {Williams}(2006)}]{rowley2006dynamics}%
  \BibitemOpen
  \bibfield  {author} {\bibinfo {author} {\bibfnamefont {C.~W.}\ \bibnamefont
  {Rowley}}\ and\ \bibinfo {author} {\bibfnamefont {D.~R.}\ \bibnamefont
  {Williams}},\ }\bibfield  {title} {\bibinfo {title} {Dynamics and control of
  high-reynolds-number flow over open cavities},\ }\href@noop {} {\bibfield
  {journal} {\bibinfo  {journal} {Annu. Rev. Fluid Mech.}\ }\textbf {\bibinfo
  {volume} {38}},\ \bibinfo {pages} {251} (\bibinfo {year} {2006})}\BibitemShut
  {NoStop}%
\bibitem [{\citenamefont {Rossiter}(1964)}]{rossiter1964wind}%
  \BibitemOpen
  \bibfield  {author} {\bibinfo {author} {\bibfnamefont {J.~E.}\ \bibnamefont
  {Rossiter}},\ }\href@noop {} {\emph {\bibinfo {title} {Wind tunnel
  experiments on the flow over rectangular cavities at subsonic and transonic
  speeds}}},\ \bibinfo {type} {Tech. Rep.}\ (\bibinfo  {institution} {Ministry
  of Aviation; Royal Aircraft Establishment; RAE Farnborough},\ \bibinfo {year}
  {1964})\BibitemShut {NoStop}%
\bibitem [{\citenamefont {Howe}(1998)}]{howe_1998}%
  \BibitemOpen
  \bibfield  {author} {\bibinfo {author} {\bibfnamefont {M.~S.}\ \bibnamefont
  {Howe}},\ }\href@noop {} {\emph {\bibinfo {title} {Acoustics of
  Fluid-Structure Interactions}}}\ (\bibinfo  {publisher} {Cambridge University
  Press},\ \bibinfo {year} {1998})\BibitemShut {NoStop}%
\bibitem [{\citenamefont {Elder}(1978)}]{elder1978self}%
  \BibitemOpen
  \bibfield  {author} {\bibinfo {author} {\bibfnamefont {S.~A.}\ \bibnamefont
  {Elder}},\ }\bibfield  {title} {\bibinfo {title} {Self-excited depth-mode
  resonance for a wall-mounted cavity in turbulent flow},\ }\href@noop {}
  {\bibfield  {journal} {\bibinfo  {journal} {J. Acoust. Soc. Am.}\ }\textbf
  {\bibinfo {volume} {64}},\ \bibinfo {pages} {877} (\bibinfo {year}
  {1978})}\BibitemShut {NoStop}%
\bibitem [{\citenamefont {Marsden}\ \emph {et~al.}(2012)\citenamefont
  {Marsden}, \citenamefont {Bailly}, \citenamefont {Bogey},\ and\ \citenamefont
  {Jondeau}}]{MARSDEN20123521}%
  \BibitemOpen
  \bibfield  {author} {\bibinfo {author} {\bibfnamefont {O.}~\bibnamefont
  {Marsden}}, \bibinfo {author} {\bibfnamefont {C.}~\bibnamefont {Bailly}},
  \bibinfo {author} {\bibfnamefont {C.}~\bibnamefont {Bogey}},\ and\ \bibinfo
  {author} {\bibfnamefont {E.}~\bibnamefont {Jondeau}},\ }\bibfield  {title}
  {\bibinfo {title} {Investigation of flow features and acoustic radiation of a
  round cavity},\ }\href
  {https://doi.org/https://doi.org/10.1016/j.jsv.2012.03.017} {\bibfield
  {journal} {\bibinfo  {journal} {J. Sound Vib.}\ }\textbf {\bibinfo {volume}
  {331}},\ \bibinfo {pages} {3521 } (\bibinfo {year} {2012})}\BibitemShut
  {NoStop}%
\bibitem [{\citenamefont {Yang}\ and\ \citenamefont
  {Morgans}(2016)}]{YANG2016294}%
  \BibitemOpen
  \bibfield  {author} {\bibinfo {author} {\bibfnamefont {D.}~\bibnamefont
  {Yang}}\ and\ \bibinfo {author} {\bibfnamefont {A.~S.}\ \bibnamefont
  {Morgans}},\ }\bibfield  {title} {\bibinfo {title} {A semi-analytical model
  for the acoustic impedance of finite length circular holes with mean flow},\
  }\href@noop {} {\bibfield  {journal} {\bibinfo  {journal} {Journal of Sound
  and Vibration}\ }\textbf {\bibinfo {volume} {384}},\ \bibinfo {pages} {294 }
  (\bibinfo {year} {2016})}\BibitemShut {NoStop}%
\bibitem [{\citenamefont {Mart{\'\i}nez-Lera}\ \emph
  {et~al.}(2009)\citenamefont {Mart{\'\i}nez-Lera}, \citenamefont {Schram},
  \citenamefont {F{\"o}ller}, \citenamefont {Kaess},\ and\ \citenamefont
  {Polifke}}]{martinez2009identification}%
  \BibitemOpen
  \bibfield  {author} {\bibinfo {author} {\bibfnamefont {P.}~\bibnamefont
  {Mart{\'\i}nez-Lera}}, \bibinfo {author} {\bibfnamefont {C.}~\bibnamefont
  {Schram}}, \bibinfo {author} {\bibfnamefont {S.}~\bibnamefont {F{\"o}ller}},
  \bibinfo {author} {\bibfnamefont {R.}~\bibnamefont {Kaess}},\ and\ \bibinfo
  {author} {\bibfnamefont {W.}~\bibnamefont {Polifke}},\ }\bibfield  {title}
  {\bibinfo {title} {Identification of the aeroacoustic response of a low mach
  number flow through a t-joint},\ }\href@noop {} {\bibfield  {journal}
  {\bibinfo  {journal} {J. Acoust. Soc. Am.}\ }\textbf {\bibinfo {volume}
  {126}},\ \bibinfo {pages} {582} (\bibinfo {year} {2009})}\BibitemShut
  {NoStop}%
\bibitem [{\citenamefont {Gikadi}\ \emph {et~al.}(2012)\citenamefont {Gikadi},
  \citenamefont {Sattelmayer},\ and\ \citenamefont {Peschiulli}}]{Gikadi2012}%
  \BibitemOpen
  \bibfield  {author} {\bibinfo {author} {\bibfnamefont {J.}~\bibnamefont
  {Gikadi}}, \bibinfo {author} {\bibfnamefont {T.}~\bibnamefont
  {Sattelmayer}},\ and\ \bibinfo {author} {\bibfnamefont {A.}~\bibnamefont
  {Peschiulli}},\ }\bibfield  {title} {\bibinfo {title} {Effects of the mean
  flow field on the thermo-acoustic stability of aero-engine combustion
  chambers},\ }in\ \href {https://doi.org/10.1115/GT2012-69612} {\emph
  {\bibinfo {booktitle} {Proceedings of the ASME Turbo Expo}}},\ Vol.~\bibinfo
  {volume} {2}\ (\bibinfo {year} {2012})\BibitemShut {NoStop}%
\bibitem [{\citenamefont {Karlsson}\ and\ \citenamefont
  {{\AA}bom}(2010)}]{karlsson2010aeroacoustics}%
  \BibitemOpen
  \bibfield  {author} {\bibinfo {author} {\bibfnamefont {M.}~\bibnamefont
  {Karlsson}}\ and\ \bibinfo {author} {\bibfnamefont {M.}~\bibnamefont
  {{\AA}bom}},\ }\bibfield  {title} {\bibinfo {title} {Aeroacoustics of
  t-junctions—an experimental investigation},\ }\href@noop {} {\bibfield
  {journal} {\bibinfo  {journal} {J. Sound Vib.}\ }\textbf {\bibinfo {volume}
  {329}},\ \bibinfo {pages} {1793} (\bibinfo {year} {2010})}\BibitemShut
  {NoStop}%
\bibitem [{\citenamefont {Fabre}\ \emph {et~al.}(2019)\citenamefont {Fabre},
  \citenamefont {Longobardi}, \citenamefont {Bonnefis},\ and\ \citenamefont
  {Luchini}}]{fabre_longobardi_bonnefis_luchini_2019}%
  \BibitemOpen
  \bibfield  {author} {\bibinfo {author} {\bibfnamefont {D.}~\bibnamefont
  {Fabre}}, \bibinfo {author} {\bibfnamefont {R.}~\bibnamefont {Longobardi}},
  \bibinfo {author} {\bibfnamefont {P.}~\bibnamefont {Bonnefis}},\ and\
  \bibinfo {author} {\bibfnamefont {P.}~\bibnamefont {Luchini}},\ }\bibfield
  {title} {\bibinfo {title} {The acoustic impedance of a laminar viscous jet
  through a thin circular aperture},\ }\href
  {https://doi.org/10.1017/jfm.2018.1008} {\bibfield  {journal} {\bibinfo
  {journal} {J. Fluid Mech.}\ }\textbf {\bibinfo {volume} {864}},\ \bibinfo
  {pages} {5–44} (\bibinfo {year} {2019})}\BibitemShut {NoStop}%
\bibitem [{\citenamefont {Fabre}\ \emph {et~al.}(2020)\citenamefont {Fabre},
  \citenamefont {Longobardi}, \citenamefont {Citro},\ and\ \citenamefont
  {Luchini}}]{fabre_longobardi_citro_luchini_2020}%
  \BibitemOpen
  \bibfield  {author} {\bibinfo {author} {\bibfnamefont {D.}~\bibnamefont
  {Fabre}}, \bibinfo {author} {\bibfnamefont {R.}~\bibnamefont {Longobardi}},
  \bibinfo {author} {\bibfnamefont {V.}~\bibnamefont {Citro}},\ and\ \bibinfo
  {author} {\bibfnamefont {P.}~\bibnamefont {Luchini}},\ }\bibfield  {title}
  {\bibinfo {title} {Acoustic impedance and hydrodynamic instability of the
  flow through a circular aperture in a thick plate},\ }\href
  {https://doi.org/10.1017/jfm.2019.953} {\bibfield  {journal} {\bibinfo
  {journal} {J. Fluid Mech.}\ }\textbf {\bibinfo {volume} {885}},\ \bibinfo
  {pages} {A11} (\bibinfo {year} {2020})}\BibitemShut {NoStop}%
\bibitem [{\citenamefont {Boujo}\ \emph {et~al.}(2018)\citenamefont {Boujo},
  \citenamefont {Bauerheim},\ and\ \citenamefont
  {Noiray}}]{boujo_bauerheim_noiray_2018}%
  \BibitemOpen
  \bibfield  {author} {\bibinfo {author} {\bibfnamefont {E.}~\bibnamefont
  {Boujo}}, \bibinfo {author} {\bibfnamefont {M.}~\bibnamefont {Bauerheim}},\
  and\ \bibinfo {author} {\bibfnamefont {N.}~\bibnamefont {Noiray}},\
  }\bibfield  {title} {\bibinfo {title} {Saturation of a turbulent mixing layer
  over a cavity: response to harmonic forcing around mean flows},\ }\href@noop
  {} {\bibfield  {journal} {\bibinfo  {journal} {J. Fluid Mech.}\ }\textbf
  {\bibinfo {volume} {853}},\ \bibinfo {pages} {386–418} (\bibinfo {year}
  {2018})}\BibitemShut {NoStop}%
\bibitem [{\citenamefont {Bauerheim}\ \emph {et~al.}(2020)\citenamefont
  {Bauerheim}, \citenamefont {Boujo},\ and\ \citenamefont
  {Noiray}}]{Bauerheim20}%
  \BibitemOpen
  \bibfield  {author} {\bibinfo {author} {\bibfnamefont {M.}~\bibnamefont
  {Bauerheim}}, \bibinfo {author} {\bibfnamefont {E.}~\bibnamefont {Boujo}},\
  and\ \bibinfo {author} {\bibfnamefont {N.}~\bibnamefont {Noiray}},\
  }\bibfield  {title} {\bibinfo {title} {Numerical analysis of the linear and
  nonlinear vortex-sound interaction in a t-junction},\ }in\ \href@noop {}
  {\emph {\bibinfo {booktitle} {AIAA AVIATION 2020 FORUM}}}\ (\bibinfo {year}
  {2020})\ p.\ \bibinfo {pages} {2569}\BibitemShut {NoStop}%
\bibitem [{\citenamefont {Bourquard}\ \emph {et~al.}(2021)\citenamefont
  {Bourquard}, \citenamefont {Faure-Beaulieu},\ and\ \citenamefont
  {Noiray}}]{bourquard_faure-beaulieu_noiray_2021}%
  \BibitemOpen
  \bibfield  {author} {\bibinfo {author} {\bibfnamefont {C.}~\bibnamefont
  {Bourquard}}, \bibinfo {author} {\bibfnamefont {A.}~\bibnamefont
  {Faure-Beaulieu}},\ and\ \bibinfo {author} {\bibfnamefont {N.}~\bibnamefont
  {Noiray}},\ }\bibfield  {title} {\bibinfo {title} {Whistling of deep cavities
  subject to turbulent grazing flow: intermittently unstable aeroacoustic
  feedback},\ }\href@noop {} {\bibfield  {journal} {\bibinfo  {journal}
  {Journal of Fluid Mechanics}\ }\textbf {\bibinfo {volume} {909}},\ \bibinfo
  {pages} {A19} (\bibinfo {year} {2021})}\BibitemShut {NoStop}%
\bibitem [{\citenamefont {Schuermans}\ \emph {et~al.}(2004)\citenamefont
  {Schuermans}, \citenamefont {Bellucci}, \citenamefont {Guethe}, \citenamefont
  {Meili}, \citenamefont {Flohr},\ and\ \citenamefont
  {Paschereit}}]{schuermans2004detailed}%
  \BibitemOpen
  \bibfield  {author} {\bibinfo {author} {\bibfnamefont {B.}~\bibnamefont
  {Schuermans}}, \bibinfo {author} {\bibfnamefont {V.}~\bibnamefont
  {Bellucci}}, \bibinfo {author} {\bibfnamefont {F.}~\bibnamefont {Guethe}},
  \bibinfo {author} {\bibfnamefont {F.}~\bibnamefont {Meili}}, \bibinfo
  {author} {\bibfnamefont {P.}~\bibnamefont {Flohr}},\ and\ \bibinfo {author}
  {\bibfnamefont {C.~O.}\ \bibnamefont {Paschereit}},\ }\bibfield  {title}
  {\bibinfo {title} {A detailed analysis of thermoacoustic interaction
  mechanisms in a turbulent premixed flame},\ }in\ \href@noop {} {\emph
  {\bibinfo {booktitle} {ASME Turbo Expo 2004: Power for Land, Sea, and Air}}}\
  (\bibinfo {organization} {American Society of Mechanical Engineers Digital
  Collection},\ \bibinfo {year} {2004})\ pp.\ \bibinfo {pages}
  {539--551}\BibitemShut {NoStop}%
\bibitem [{\citenamefont {Kooijman}\ \emph {et~al.}(2008)\citenamefont
  {Kooijman}, \citenamefont {Hirschberg},\ and\ \citenamefont
  {Golliard}}]{KOOIJMAN2008849}%
  \BibitemOpen
  \bibfield  {author} {\bibinfo {author} {\bibfnamefont {G.}~\bibnamefont
  {Kooijman}}, \bibinfo {author} {\bibfnamefont {A.}~\bibnamefont
  {Hirschberg}},\ and\ \bibinfo {author} {\bibfnamefont {J.}~\bibnamefont
  {Golliard}},\ }\bibfield  {title} {\bibinfo {title} {Acoustical response of
  orifices under grazing flow: Effect of boundary layer profile and edge
  geometry},\ }\href
  {https://doi.org/https://doi.org/10.1016/j.jsv.2008.02.030} {\bibfield
  {journal} {\bibinfo  {journal} {Journal of Sound and Vibration}\ }\textbf
  {\bibinfo {volume} {315}},\ \bibinfo {pages} {849 } (\bibinfo {year}
  {2008})}\BibitemShut {NoStop}%
\bibitem [{\citenamefont {Howe}\ \emph {et~al.}(1996)\citenamefont {Howe},
  \citenamefont {Scott},\ and\ \citenamefont {Sipcic}}]{Howe1996}%
  \BibitemOpen
  \bibfield  {author} {\bibinfo {author} {\bibfnamefont {M.~S.}\ \bibnamefont
  {Howe}}, \bibinfo {author} {\bibfnamefont {M.~I.}\ \bibnamefont {Scott}},\
  and\ \bibinfo {author} {\bibfnamefont {S.~R.}\ \bibnamefont {Sipcic}},\
  }\bibfield  {title} {\bibinfo {title} {The influence of tangential mean flow
  on the rayleigh conductivity of an aperture},\ }\href@noop {} {\bibfield
  {journal} {\bibinfo  {journal} {Proc. R. Soc. London A}\ }\textbf {\bibinfo
  {volume} {452}},\ \bibinfo {pages} {2303} (\bibinfo {year}
  {1996})}\BibitemShut {NoStop}%
\bibitem [{\citenamefont {Hirschberg}(2007)}]{hirschberg2007introduction}%
  \BibitemOpen
  \bibfield  {author} {\bibinfo {author} {\bibfnamefont {A.}~\bibnamefont
  {Hirschberg}},\ }\bibfield  {title} {\bibinfo {title} {Introduction to
  aero-acoustics of internal flows},\ }in\ \href@noop {} {\emph {\bibinfo
  {booktitle} {Basics of aeroacoustics and thermoacoustics}}}\ (\bibinfo
  {publisher} {Von Karman Institute for Fluid Dynamics},\ \bibinfo {year}
  {2007})\ pp.\ \bibinfo {pages} {1--112}\BibitemShut {NoStop}%
\bibitem [{\citenamefont {Debnath}\ and\ \citenamefont
  {Bhatta}(2014)}]{debnath2014integral}%
  \BibitemOpen
  \bibfield  {author} {\bibinfo {author} {\bibfnamefont {L.}~\bibnamefont
  {Debnath}}\ and\ \bibinfo {author} {\bibfnamefont {D.}~\bibnamefont
  {Bhatta}},\ }\href@noop {} {\emph {\bibinfo {title} {Integral transforms and
  their applications}}}\ (\bibinfo  {publisher} {CRC press},\ \bibinfo {year}
  {2014})\BibitemShut {NoStop}%
\bibitem [{\citenamefont {Howe}(1997)}]{Howe97}%
  \BibitemOpen
  \bibfield  {author} {\bibinfo {author} {\bibfnamefont {M.~S.}\ \bibnamefont
  {Howe}},\ }\bibfield  {title} {\bibinfo {title} {Low strouhal number
  instabilities of flow over apertures and wall cavities},\ }\href@noop {}
  {\bibfield  {journal} {\bibinfo  {journal} {J. Acoust. Soc. Am.}\ }\textbf
  {\bibinfo {volume} {102}},\ \bibinfo {pages} {772} (\bibinfo {year}
  {1997})}\BibitemShut {NoStop}%
\bibitem [{\citenamefont {Howe}(1979)}]{howe1979theory}%
  \BibitemOpen
  \bibfield  {author} {\bibinfo {author} {\bibfnamefont {M.~S.}\ \bibnamefont
  {Howe}},\ }\bibfield  {title} {\bibinfo {title} {On the theory of unsteady
  high reynolds number flow through a circular aperture},\ }\href@noop {}
  {\bibfield  {journal} {\bibinfo  {journal} {Proc. R. Soc. London A}\ }\textbf
  {\bibinfo {volume} {366}},\ \bibinfo {pages} {205} (\bibinfo {year}
  {1979})}\BibitemShut {NoStop}%
\bibitem [{\citenamefont {Howe}(1981)}]{howe_1981}%
  \BibitemOpen
  \bibfield  {author} {\bibinfo {author} {\bibfnamefont {M.~S.}\ \bibnamefont
  {Howe}},\ }\bibfield  {title} {\bibinfo {title} {The influence of mean shear
  on unsteady aperture flow, with application to acoustical diffraction and
  self-sustained cavity oscillations},\ }\href
  {https://doi.org/10.1017/S0022112081000979} {\bibfield  {journal} {\bibinfo
  {journal} {J. Fluid Mech.}\ }\textbf {\bibinfo {volume} {109}},\ \bibinfo
  {pages} {125–146} (\bibinfo {year} {1981})}\BibitemShut {NoStop}%
\bibitem [{\citenamefont {Takahashi}\ \emph {et~al.}(2016)\citenamefont
  {Takahashi}, \citenamefont {Iwagami}, \citenamefont {Kobayashi},\ and\
  \citenamefont {Takami}}]{takahashi2016theoretical}%
  \BibitemOpen
  \bibfield  {author} {\bibinfo {author} {\bibfnamefont {K.}~\bibnamefont
  {Takahashi}}, \bibinfo {author} {\bibfnamefont {S.}~\bibnamefont {Iwagami}},
  \bibinfo {author} {\bibfnamefont {T.}~\bibnamefont {Kobayashi}},\ and\
  \bibinfo {author} {\bibfnamefont {T.}~\bibnamefont {Takami}},\ }\bibfield
  {title} {\bibinfo {title} {Theoretical estimation of the acoustic energy
  generation and absorption caused by jet oscillation},\ }\href@noop {}
  {\bibfield  {journal} {\bibinfo  {journal} {J. Phys. Soc. Jpn.}\ }\textbf
  {\bibinfo {volume} {85}},\ \bibinfo {pages} {044402} (\bibinfo {year}
  {2016})}\BibitemShut {NoStop}%
\bibitem [{\citenamefont {Abramowitz}\ and\ \citenamefont
  {Stegun}(1948)}]{abramowitz1948handbook}%
  \BibitemOpen
  \bibfield  {author} {\bibinfo {author} {\bibfnamefont {M.}~\bibnamefont
  {Abramowitz}}\ and\ \bibinfo {author} {\bibfnamefont {I.~A.}\ \bibnamefont
  {Stegun}},\ }\href@noop {} {\emph {\bibinfo {title} {Handbook of mathematical
  functions with formulas, graphs, and mathematical tables}}},\ Vol.~\bibinfo
  {volume} {55}\ (\bibinfo  {publisher} {US Government printing office},\
  \bibinfo {year} {1948})\BibitemShut {NoStop}%
\bibitem [{\citenamefont {{Hansen}}\ \emph {et~al.}(2009)\citenamefont
  {{Hansen}}, \citenamefont {{Niederberger}}, \citenamefont {{Guzzella}},\ and\
  \citenamefont {{Koumoutsakos}}}]{CMAES}%
  \BibitemOpen
  \bibfield  {author} {\bibinfo {author} {\bibfnamefont {N.}~\bibnamefont
  {{Hansen}}}, \bibinfo {author} {\bibfnamefont {A.~S.~P.}\ \bibnamefont
  {{Niederberger}}}, \bibinfo {author} {\bibfnamefont {L.}~\bibnamefont
  {{Guzzella}}},\ and\ \bibinfo {author} {\bibfnamefont {P.}~\bibnamefont
  {{Koumoutsakos}}},\ }\bibfield  {title} {\bibinfo {title} {A method for
  handling uncertainty in evolutionary optimization with an application to
  feedback control of combustion},\ }\href@noop {} {\bibfield  {journal}
  {\bibinfo  {journal} {IEEE Trans. Evolut. Comput.}\ }\textbf {\bibinfo
  {volume} {13}},\ \bibinfo {pages} {180} (\bibinfo {year} {2009})}\BibitemShut
  {NoStop}%
\end{thebibliography}%

\end{document}